\documentclass[%
 aip,
 amsmath,amssymb,
preprint,
reprint, onecolumn 
]{revtex4-1}
\usepackage{amsfonts}
\usepackage{amsmath}
\usepackage{amssymb}
\usepackage{amsthm}
\usepackage{stmaryrd}
\usepackage{physics}
\usepackage{braket}
\usepackage{dsfont}
\usepackage{bbold}
\usepackage{graphicx}
\usepackage{relsize}
\usepackage{centernot}
\usepackage{makecell}
\usepackage[table,xcdraw]{xcolor}
\usepackage{enumerate}
\usepackage[pagebackref, colorlinks = true, linkcolor = blue, urlcolor  = blue, citecolor = red]{hyperref}
\usepackage[margin=1in]{geometry}
\usepackage{enumitem}
\usepackage{mathtools}
\usepackage{comment}
\usepackage{float}
\usepackage[capitalize]{cleveref}
\usepackage{booktabs}
\usepackage{multirow}
\usepackage[skins]{tcolorbox}
\usepackage{nicematrix}
\NiceMatrixOptions{cell-space-limits = 2pt}
\usetikzlibrary{matrix, arrows.meta}
\usepackage{fancyhdr}
\usepackage{etoolbox}
\usepackage{standalone}
\usepackage{adjustbox}

\usetikzlibrary{graphs,graphs.standard}

\def\lf{\left\lfloor}   
\def\rf{\right\rfloor}

\newcommand{\CircPhase}[1]{\mathfrak{C}\mathcal{DU}(#1)}
\newcommand{\CircSign}[1]{\mathfrak{C}\mathcal{DO}(#1)}

\DeclareMathOperator{\ext}{ext}
\newcommand{\PPTcone}{\mathsf{PPT}}

\newcommand{\SEP}[1]{\mathsf{SEP}_{#1}}
\newcommand{\SEPcone}{\mathsf{SEP}}
\newcommand{\PSD}[1]{\mathsf{PSD}_{#1}}
\newcommand{\circEWP}[1]{\overrightarrow{\mathsf{EWP}}_{#1}}
\newcommand{\circPSD}[1]{\overrightarrow{\mathsf{PSD}}_{#1}}
\newcommand{\circCP}[1]{\overrightarrow{\mathsf{CP}}_{#1}}

\newcommand{\circTCP}[1]{\overrightarrow{\mathsf{TCP}}_{#1}}
\newcommand{\COP}[1]{\mathsf{COP}_{#1}}
\newcommand{\circDNN}[1]{\overrightarrow{\mathsf{DNN}}_{#1}}
\newcommand{\circCOP}[1]{\overrightarrow{\mathsf{COP}}_{#1}}
\newcommand{\circSPN}[1]{\overrightarrow{\mathsf{SPN}}_{#1}}

\newcommand{\CP}[1]{\mathsf{CP}_{#1}}
\newcommand{\SPN}[1]{\mathsf{SPN}_{#1}}
\newcommand{\DNN}[1]{\mathsf{DNN}_{#1}}

\newcommand{\LDOIcone}{\mathsf{LDOI}}

\newcommand{\LCSIcone}{\mathsf{LCSI}}
\newcommand{\LCSI}[1]{\mathsf{LCSI}_{#1}}
\newcommand{\LDOI}[1]{\mathsf{LDOI}_{#1}}

\newcommand{\EWP}[1]{\mathsf{EWP}_{#1}}
\newcommand{\rev}{\mathsf{R}}
\newcommand{\cone}{\operatorname{cone}}
\newcommand{\circul}[1]{\mathsf{circ}(#1)}
\newcommand{\circulinv}[1]{\mathsf{circ}^{-1}(#1)}

\newcommand{\circulant}[2]{\mathsf{C}_#2^{#1}}

\renewcommand{\epsilon}{\varepsilon}
\renewcommand{\phi}{\varphi}
\newcommand{\overbar}[1]{\mkern 1.5mu\overline{\mkern-1.5mu#1\mkern-1.5mu}\mkern 1.5mu}

\newcommand{\MLDUI}[1]{\mathcal{M}_{#1}
\newcommand{\rom}[1]{\expandafter\@slowromancap\romannumeral #1@}
(\mathbb{C})^{\times 2}_{\mathbb{C}^{#1}}}
\newcommand{\MLDOI}[1]{\mathcal{M}_{#1}(\mathbb{C})^{\times 3}_{\mathbb{C}^{#1}}}

\newcommand{\C}{\mathbb{C}}
\renewcommand{\C}[1]{\mathbb{C}^{#1}}
\newcommand{\M}[1]{\mathcal{M}_{#1}(\mathbb{C})}

\newcommand{\Mreal}[1]{\mathcal{M}_{#1}(\mathbb{R})}
\newcommand{\Msareal}[1]{\mathcal{M}_{#1}^{sa}(\mathbb{R})}
 
\newcommand{\U}[1]{\mathcal{U}(#1)}
\newcommand{\id}[1]{\mathrm{I}_{n}}

\newcommand{\Msa}[1]{\mathcal{M}^{sa}_{#1}(\mathbb{C})}

\newcommand{\OO}[1]{(#1\otimes#1)}

\newcommand{\R}[1]{\mathbb{R}^{#1}}

\newcommand{\channel}{\mathrm{\Phi}}

\DeclareMathOperator{\diag}{diag}

\DeclareMathOperator{\supp}{supp}

\newtheorem{theorem}{Theorem}[section]
\newtheorem{definition}[theorem]{Definition}

\newtheorem{proposition}[theorem]{Proposition}
\newtheorem{corollary}[theorem]{Corollary}
\newtheorem{lemma}[theorem]{Lemma}
\newtheorem{remark}[theorem]{Remark}

\newtheorem{example}[theorem]{Example} 

\begin{document}

\title{Entanglement in cyclic sign invariant quantum states}

\author{Aabhas Gulati}
\email{aabhas.gulati@math.univ-toulouse.fr}
\address{Institut de Mathématiques, Université de Toulouse, UPS, France.}

\author{Ion Nechita}
\email{ion.nechita@univ-tlse3.fr}
\address{Laboratoire de Physique Th\'eorique, Universit\'e de Toulouse, CNRS, UPS, France.}

\author{Satvik Singh}
\email{ss2821@cam.ac.uk}
\address{\parbox{\linewidth}{Department of Applied Mathematics and Theoretical Physics, \\ University of Cambridge, Cambridge, United Kingdom.}}

\begin{abstract}
    We introduce and study bipartite quantum states that are invariant under the local action of the cyclic sign group. Due to symmetry, these states are sparse and can be parameterized by a triple of vectors. Their important semi-definite properties, such as positivity and positivity under partial transpose (PPT), can be simply characterized in terms of these vectors and their discrete Fourier transforms. We study in detail the entanglement properties of this family of symmetric states, showing that it contains PPT entangled states. For states that are diagonal in the Dicke basis, deciding separability is equivalent to a circulant version of the complete positivity problem. In local dimension $d\leq 5$, we completely characterize these sets and construct entanglement witnesses; some partial results are also obtained for $d=6,7$. We construct a new family of states for which the properties of PPT and separability can be characterized for all dimensions, generalizing results from from the literature. Our results show that these states have a rich entanglement structure, even in the bosonic subspace.
\end{abstract}

\maketitle
\tableofcontents

\section{Introduction}

With the significant advent of quantum information science over the past few decades, entanglement has emerged as a fundamental resource, with practical applications in cryptography \cite{Ekert1991crypto}, communication \cite{Bennett1993teleport}, and computation \cite{Raussendorf2001computer}. However, entanglement theory has inherent computational complexity: it is NP-hard to decide whether a given bipartite quantum state is entangled or not \cite{gurvits2003classical}. Consequently, several sufficient conditions to detect the presence of entanglement have been developed \cite{Horodecki2009entanglement}. Among them, perhaps the simplest one is the so-called positivity under partial transpose (PPT) criterion \cite{peres1996separability}: any unentangled (or separable) bipartite state remains positive after transposition is applied to either one of the subfactors. In low dimensions, this criterion is necessary and sufficient to detect entanglement: any qubit-qubit or qubit-qutrit quantum state is separable if and only if it is PPT \cite{Strmer1963, horodecki1996separability}. However, in higher dimensions, there exist quantum states that are PPT but still entangled \cite{Horodecki1997PPTent}. Curiously, one cannot distill any pure entanglement from such PPT entangled states by using local operations and classical communication (LOCC) \cite{Horodecki1998bound}, even though the entanglement cost of preparing such states under LOCC is non-zero \cite{Vidal2001irreversible, Yang2005irreversible}. Consequently, the existence of PPT entangled states is closely related to the irreversibility of the resource theory of entanglement \cite{Vidal2001irreversible, Lami2023irreversible}.

Another perspective on the complexity of entanglement is of a geometric nature. The sets of separable and PPT states in a $d\otimes d$ system are convex bodies in a high-dimensional ($d^4-1$) space. Consequently, these sets are difficult to characterize geometrically. One way to tackle this problem is by imposing symmetries to reduce the dimension of the state space, which makes the problem more tractable \cite{Werner1989, Horodecki1999iso, vollbrecht2001entanglement}. The maximal (local) symmetry one can impose on a $d\otimes d$ system is that of invariance under the full unitary group $\U{d}$: 
$$ \forall U\in \U{d}: \quad \OO{U} \rho \OO{U}^* = \rho \quad \text{   or   } \quad  (U \otimes \bar U) \rho (U \otimes \bar U)^* = \rho .$$ 
Quantum states with this symmetry are known as \emph{Werner} \cite{Werner1989} and \emph{isotropic} \cite{Horodecki1999iso} states, respectively. These states can be described by a single real parameter. Crucially, the convex sets of PPT and separable states coincide under this symmetry and can be characterized completely (see \cite[Fig 1]{vollbrecht2001entanglement}). In other words, the constraint of full local unitary invariance is too severe to accommodate the existence of PPT entanglement. Similar results also hold for the class of states that are invariant under the full orthogonal group $\mathcal{O}(d)$ \cite{vollbrecht2001entanglement, keyl2002fundamentals}. It is then natural to relax the symmetry constraints by considering subgroups $G\subseteq \U{d}$ and $H\subseteq \mathcal{O}(d)$ that are large enough to keep the dimension of the corresponding invariant spaces tractable, but not too large, lest the problem becomes trivial as above.

In recent work \cite{nechita2021graphical,singh2021diagonal,Singh2020entanglement}, some of the authors of this work have considered the subgroups of diagonal unitary and diagonal orthogonal matrices: $G=D\mathcal{U}(d)$ and $H=D\mathcal{O}(d)$, respectively. States that are invariant under the local action of these groups (called Local Diagonal Unitary/Orthogonal Invariant or LDUI/LDOI) can be parametrized by a triple of $d\times d$ matrices $(A,B,C)$ with a common diagonal, and the convex sets of PPT and separable states can be described by imposing suitable positivity conditions on these matrices. Except for $d=2$, PPT is not equivalent to separability, and many examples of PPT entangled states can be constructed with this symmetry \cite{Singh2020entanglement}. A well known class of LDOI states are mixtures of the so-called \emph{Dicke} states \cite{yu2016separability} \cite{tura2018separability}, which correspond to matrix triples of the form $(A,\diag(A),A)$ in the LDOI parametrization. The problem of detecting entanglement for these states is equivalent to the well known complete positivity problem from optimization theory \cite{berman2003completely}, which is also known to be NP-hard \cite{tura2018separability}.

\subsection{Summary of results} In this paper, we interpolate between the full unitary/orthogonal groups and their diagonal counterparts by considering the \emph{cyclic phase} and \emph{cyclic sign} subgroups: $G=\CircPhase{d}:= \operatorname{Cyc}(d) \ltimes \mathcal{DU}(d)$ and  $H=\CircSign{d}:= \operatorname{Cyc}(d) \ltimes \mathcal{DO}(d)$, respectively. Here, $\operatorname{Cyc}(d)$ is the abelian group of \emph{cyclic permutations} of $d$ elements and $\ltimes$ denotes the semi-direct product. States that are locally invariant under these groups are called \emph{Local Cyclic Phase Invariant} (LCPI) and \emph{Local Cyclic Sign Invariant} (LCSI), respectively. These states lie in between the classes discussed above, satisfying reverse inclusion relations to those satisfied by the symmetry groups (see Eqs~\eqref{eq:LCPI},\eqref{eq:LCSI}). The cyclic symmetry forces the $(A,B,C)$ matrices in the LDOI parameterization of these states to be \emph{circulant}. Consequently, these states can be parameterized by a triple of vectors $(a,b,c)$ with a common first entry, allowing them to be expressed by $3d-2$ parameters for a given local dimension $d$. Despite the highly symmetric form of these states, we will see that PPT entanglement still exists in this class for all $d\geq 3$ [See \cref{sec:examples}]

\begin{tcolorbox}[
    colback=blue!3,
    colframe=black!50,
    boxrule=0.5mm,
    rounded corners,
    width=\textwidth,
]
\vspace{3pt}
\begin{align}
\label{eq:LCPI} D\mathcal{U}(d) \, \, (\text{LDUI states}) \leq {\CircPhase{d}} \, \, &(\textbf{LCPI states}) \leq \mathcal{U}(d) \, \,(\text{Werner/isotropic states})\\[15pt]
\label{eq:LCSI}D\mathcal{O}(d) \, \, (\text{LDOI states}) \leq {\CircSign{d}} \, \, &(\textbf{LCSI states}) \leq \mathcal{O}(d) \, \,(\text{Brauer states})
\end{align}
\vspace{0pt}
\end{tcolorbox} 

We study the convex structure of several cones with cyclic symmetry that are relevant from the perspective of quantum information, such as the cones of positive semidefinite, PPT, and separable matrices. Each of these cones can be characterized simply by imposing some positivity conditions on the vector triples $(a,b,c)$. In particular, we introduce the notion of \emph{Circulant} Triplewise Completely Positive vector triples that characterizes separability for LCPI/LCSI states. This is inspired by the results in \cite{singh2021diagonal}.  Crucially, the circulant structure makes the \emph{Fourier transform} a critical tool to analyze the positivity properties of LCPI/LCSI states.

We study mixtures of Dicke states with cyclic symmetry in detail. We analyze PPT and separable states of this form, providing explicit results in low dimensions and emphasizing PPT entanglement when present. The problem of determining separability or PPT in this class reduce to a circulant version of deciding if a given matrix is completely positive or doubly non-negative, respectively \cite{tura2018separability} \cite{berman2003completely}. For all $d \geq 5$, we prove the existence of PPT entangled states in this class. Importantly, the PPT cone restricted to this family is \emph{polyhedral}; we enumerate the extremal rays in small dimension ($d \leq 7$) and relate the general case to that of computing the \emph{semi-positive cone} \cite{sivakumar2018semipositive} of the Fourier matrix. We explicitly compute \emph{all} the extremal rays of the separable cone and its dual in the first non-trivial case ($d=5$) and provide some partial results for $d=6,7$, showing the presence of PPT entanglement.

In \cref{tab:families}, we present all the classes of symmetric states introduced above, emphasizing the existence of PPT entanglement; see \cref{sec:families-symmetric-states} for more details.  
Let us mention here that circulant symmetry has received some attention in the literature, in particular with respect to the \emph{Quantum Fourier Transform} \cite{zhou2017quantum,ivanov2020two,yachi2023implementing}.

\begin{table}
\bgroup
\def\arraystretch{1.5}

\begin{tabular}{|r|l|c|c|c|c|}
\hline
\rowcolor[HTML]{C0C0C0} 
Group                                           & Inv. Q. States                                                                & Dim. Inv. & Abelian & $\exists$ PPT ent. & References                                             \\ \hline
$\{\mathrm{id}\}$                                            & all states                                                                    & $d^4$          & Y       & Y                  & -                                                      \\ \hline
$\mathcal{DO}(d)$                               & LDOI                                                                          & $3d^2-2d$      & Y       & Y                  &                                                        \\ \cline{1-5}
$\mathcal{DU}(d)$                               & (C)LDUI                                                                       & $2d^2-d$       & Y       & Y                  & \multirow{-2}{*}{\cite{singh2021diagonal}}             \\ \hline
\rowcolor[HTML]{ECF4FF} 
$\operatorname{Cyc(d)} \ltimes \mathcal{DO}(d)$ & LCSI                                                                          & $3d-2$         & N       & Y                  &                                                        \\ \cline{1-5}
\rowcolor[HTML]{ECF4FF} 
$\operatorname{Cyc(d)} \ltimes \mathcal{DU}(d)$ & LCPI                                                                          & $2d-1$         & N       & Y                  & \multirow{-2}{*}{this paper}                           \\ \hline
$\operatorname{Sym(d)} \ltimes \mathcal{DO}(d)$ & hyperoctahedral                                                              & $4$            & N       & N                  & \cite{park2024universal}                               \\ \hline
$\mathcal{O}(d)$                                & Brauer                                                                        & $3$            & N       & N                  & \cite{vollbrecht2001entanglement,keyl2002fundamentals} \\ \hline
$\mathcal{U}(d)$                                & \begin{tabular}[c]{@{}l@{}}Werner ($UU$)\\ isotropic ($U\bar U$)\end{tabular} & $2$            & N       & N                  & \cite{Werner1989,Horodecki1999iso}                     \\ \hline
\end{tabular}

\egroup
\caption{Families of symmetric bipartite quantum states and their invariance groups. In the third column we list the dimension of the (real) vector space of invariant bipartite self-adjoint \emph{matrices}. The fourth and fifth columns list the commutativity property of the invariance group and respectively whether the family of invariant matrices contains PPT entangled states.}
\label{tab:families}
\end{table}

\subsection{Outline of the paper} We provide some background on the separability problem, convex geometry and circulant matrices in \cref{sec:prelims}. In \cref{sec:families-symmetric-states}, we describe some known families of symmetric states, as presented in \cref{tab:families}. In \cref{sec:LCSI}, we introduce the families of LCPI/LCSI states and explore some of their basic properties. The linear and convex structure of LCPI/LCSI states is explored in \cref{sec:linear-space} and \cref{sec:convex-lcsi}, respectively. \cref{sec:cyclic-Dicke} contains results about an important subclass of LCSI states: \emph{cyclic mixture of Dicke states}. In \cref{sec:examples}, we introduce a new class of cyclic sign invariant states for which the separability can be completely characterized.

\section{Preliminaries}
\label{sec:prelims}
\subsection{Notation}
We start by defining the notation used throughout this paper. A vector $v$ is an element in either \(\mathbb{C}^d\) or \(\mathbb{R}^d\), and is labelled using vector components starting from index \(0\). We sometimes also use Dirac's \emph{bra-ket} notation to write vectors. In this notation, column vectors $v \in \mathbb{C}^d$ are written as kets $\ket{v}$ and their dual row vectors (conjugate transposes) $v^* \in (\mathbb{C}^d)^*$ are written as bras $\bra{v}$. The standard \emph{inner product} $v^*w= \braket{v,w}$ on $\mathbb{C}^d$ is denoted by $\langle v | w \rangle$ and the rank one matrix $vw^*$ is denoted by the \emph{outer product} $\ketbra{v}{w}$. The standard basis in $\C{d}$ is denoted by $\{ \ket{i}\}_{i\in [d]}$, where $[d]:= \{0,1,\ldots ,d-1 \}$.

We define $\M{d}$ as the set of \(d \times d\) complex matrices and $\Msa{d}:= \{A\in \M{d} : A=A^* \}$ as the set of $d\times d$ self-adjoint complex matrices, where the conjugate transpose of $A\in \M{d}$ is denoted by $A^*$. $\Mreal{d}$ and $\Msareal{d}$ are defined similarly for real matrices. The cone of positive semi-definite matrices in $\M{d}$ is denoted by $\PSD{d}$, and the cone of entry-wise non-negative matrices by $\EWP{d}$. The set of all linear maps $\Phi : \M{d}\to \M{d}$ is denoted by \(\mathcal{T}_d(\mathbb{C})\). A map \(\mathcal{E} \in \mathcal{T}_d(\mathbb{C})\) is called \textit{positive} if \(\mathcal{E}(X) \in \PSD{d}\) for all \(X \in \PSD{d}\). We say that a map \(\mathcal{E}\) is \textit{\(k\)-positive} if the map \(\operatorname{id}_n \otimes \, \mathcal{E} : \mathcal{M}_n \otimes \mathcal{M}_d \rightarrow \mathcal{M}_n \otimes \mathcal{M}_d\) is positive for all \(1 \leq n \leq k\), where $\operatorname{id}_n:\M{n}\to \M{n}$ is the identity map. A map that is \(k\)-positive for all \(k\in \mathbb{N}\) is called \textit{completely positive}. The linear transposition map \(\text{T} : \M{d} \rightarrow \M{d}\) is positive but not 2-positive. Finally, we denote by \(\mathcal{F}: \mathbb{C}^d \rightarrow \mathbb{C}^d\) the discrete Fourier transform, which maps 
\[
x \mapsto \mathcal{F}x = \left( \frac{1}{\sqrt{d}} \sum_{j=0}^{d-1} x_j \omega^{jk} \right)^{d-1}_{k=0}
\]
where \(\omega = e^{2 \pi \mathrm i / d}\) is a primitive \(d\)-th root of unity.

\subsection{Separability and PPT entanglement}

\begin{definition}
    A bipartite positive matrix $\rho \in \M{d} \otimes \M{d}$ is said to be separable 
    if $$\rho = \sum^N_{i=1}\ketbra{v_i}{v_i} \otimes \ketbra{w_i}{w_i}$$ for some finite set of vectors $\ket{v_i}, \ket{w_i} \in \C{d}$, and it is said to be entangled otherwise.
\end{definition}

We denote the convex cone of all separable matrices in $\M{d}\otimes \M{d}$ by $\SEPcone_d$. By the Hahn-Banach hyperplane theorem, it is possible to separate this set from every entangled state using a hyperplane. For any \textit{entangled state} $\rho$ we can find a Hermitian operator $W$ such that 

\begin{itemize}
    \item $\text{Tr}(\sigma W) \geq 0$ for all $\sigma$ in  $\SEP{d}$
 \item $\text{Tr}(\rho W) < 0$ 
\end{itemize} 

This Hermitian operator is called an entanglement witness. Horodecki's criterion \cite{horodecki1996separability} gives us a operational way to detect entanglement, by finding a positive map $\channel$ such that  $(\operatorname{id}_d \otimes \channel) (\rho)$ is not positive semi-definite. One important positive map in this regard is the transposition map $\text{T}$. The states which are not positive under transposition are entangled, while the rest of the states are called $\PPTcone$ states, which also includes the set of $\SEPcone$ states.  This criterion to verify entanglement is called the $\text{PPT}$ (Positivity under Partial Transpose) criterion. In the case of $d = 2$, the reverse implication is also true, i.e. any state that is $\PPTcone$ is also separable \cite{Strmer1963, horodecki1996separability}. This is not true for $d \geq 3$. The states that satisfy the $\PPTcone$ condition but are still entangled (for $d \geq 3$) are called $\PPTcone$ entangled states \cite{Horodecki1997PPTent}.

\begin{definition}[Separability Problem]
Given a bipartite density matrix $\rho \in \M{d} \otimes \M{d}$, decide whether $\rho \in$  $\SEPcone$ or not. 
\end{definition}

It is well known that the membership problem (and the weak membership problem) for $\SEPcone$ is NP-hard \cite{gurvits2003classical, gharibian2010strong}. Unless P = NP, there is no computationally efficient criterion to decide if a state is separable or entangled. In later sections, we will study the $\SEPcone$ problem for the class of symmetric states. 

\subsection{Circulant Matrices}
Circulant matrices are highly symmetric matrices that appear naturally in many areas of mathematics \cite{davis1979circulant}. We start with their basic definition. 
\begin{definition}
    The \emph{circulant} matrix $A\in \M{d}$ associated with a vector $a\in \C{d}$, denoted $ A= \circul{a}$, is defined entrywise as follows
\[ A_{ij} = a_{(j-i) \, \operatorname{mod} \, d}. \]
A matrix $A\in \M{d}$ is said to be \emph{circulant} if $A=\circul{a}$ for some $a\in \C{d}$. We denote the set of all circulant matrices in $\M{d}$ by $\mathsf{Circ}_d$.
\end{definition}

 If we define the right cyclic shift $T : \C{d} \rightarrow \C{d}$ as $S: (a_0, a_1, \ldots a_{d-1}) \rightarrow (a_{d-1}, a_0, \ldots a_{d-2})$, it is clear that the rows of the matrix $A=\circul{a}$ are $a, S(a), S^2 (a) \ldots S^{d-1} (a)$: 

\[
A = \begin{pmatrix}
a_0 & a_1 & a_2 & \cdots & a_{d-2} & a_{d-1} \\
a_{d-1} & a_0 & a_1 & \cdots & a_{d-3} & a_{d-2} \\
a_{d-2} & a_{d-1} & a_0 & \cdots & a_{d-2} & a_{d-3} \\
\vdots & \vdots & \vdots & \ddots & \vdots & \vdots \\
a_2 & a_3 & a_4 & \cdots & a_0 & a_1 \\
a_1 & a_2 & a_3 & \cdots & a_{d-1} & a_0
\end{pmatrix}.
\]

\begin{remark}\label{Remark:shift}
    $\mathsf{Circ}_d \subseteq \M{d}$ forms a $d$-dimensional commutative algebra with the standard operations of matrix addition and matrix multiplication. Recall that $\ket 1 = (0\, 1 \, 0\, \cdots \, 0)^\top$ is the second canonical basis vector and let $P = \circul{\ket{1}} \in \M{d}$ be a shift permutation. Then, any circulant matrix $A\in \M{d}$ can be written as \[A = \sum_{i=0}^{d-1} a_i P^i, \quad \text{where } a_i = A_{0,i}.  \]  
\end{remark}

Recall that the Fourier transform matrix $\mathcal{F}\in \M{d}$ is defined entrywise as $\mathcal{F}_{jk}=\omega^{jk}/\sqrt{d}$, where $\omega=e^{2\pi \mathrm{i}/d}$ is the $d^{\text{th}}$ primitive root of unity. The $1/\sqrt{d}$ factor ensures that $\mathcal{F}$ is unitary. The inverse Fourier transform $\mathcal{F}^{-1}=\mathcal{F}^{*}$ is given by $(\mathcal{F}^{-1})_{jk} = \omega^{-jk}/\sqrt{d}$.

\begin{proposition}[{{\cite[Chapter 3.2]{davis1979circulant}}}]
    The eigenvalues $\{\lambda_i\}_{i=0}^{d-1}$ of a circulant matrix $A  = \circul{a} \in \M{d}$ are obtained by taking the Fourier transform of $a$:
    $$\forall j \in [d], \qquad \lambda_j = (\sqrt{d} \mathcal{F}a)_j  = \sum_k a_k \omega^{jk},$$
    where $\omega=e^{2\pi i/d}$ is the $d^{\text{th}}$ primitive root of unity.
\end{proposition}
\begin{proof}
    This follows from the fact that the Fourier matrix diagonalizes circulant matrices: 
    $$\mathcal F^* \cdot \operatorname{circ}(a) \cdot \mathcal F = \operatorname{diag}(\sqrt d \mathcal F a)$$
\end{proof}

\begin{remark}
    A circulant matrix $A=\circul{a}\in \M{d}$ is Hermitian iff $a_i = \overbar{a_i}^\rev = \overbar{a_{d-i}}$. The \emph{reversal} operation $a \rightarrow a^\rev$, is defined as 
    $$\forall i \in [d] \qquad (a^\rev)_i := a_{(-i) \operatorname{mod} d}.$$ In the case of circulant matrices, we have: $\circul{a^\rev} = \circul{a}^\top$.
\end{remark}

\begin{definition}
    The bilinear circular convolution map $* : \C{d} \times \C{d} \rightarrow \C{d}$ is defined as 
    $$\forall k \in [d], \qquad  {(a * b)}_k  = \sum_{j=0}^{d-1} a_j b_{k - j} = \bra{\overbar{a}} P^{-k} \ket{b^\rev} $$
\end{definition}

The circular convolution corresponds exactly to the multiplication of two circulant matrices: 
$$\forall a,b\in \C{d}: \qquad \circul{a} \circul{b} = \circul{a * b}=\circul{b * a} = \circul{b} \circul{a}.$$

Finally, the Hilbert-Schmidt inner product of circulant matrices corresponds to the euclidean inner product of vectors, 
$$\operatorname{Tr}(\circul{a}^* \circul{b}) =  d \braket{a | b} $$

\subsection{Convex cones and extremal rays}
In this section, we introduce some basic notions from convex analysis.
\begin{definition}
    Let $V$ be a real vector space. A \emph{convex cone} $\mathcal C$ is a subset of $V$ having the following two properties: 
    \begin{itemize}
        \item if $x \in \mathcal C$ and $\lambda \in \mathbb R^+ = [0, \infty)$, then $\lambda x \in \mathcal C$.
        \item if $x,y \in \mathcal C$, then $x+y \in \mathcal C$.
    \end{itemize}
    In particular, $0 \in \mathcal C$. The cone $\mathcal C$ is said to be \emph{pointed} if $\mathcal C \cap (-\mathcal C) = \{0\}$; in other words, $\mathcal C$ is pointed if it does not contain any line. 
    
    For a vector $v \neq 0$, the half-line $\mathbb R_+ v := \{\lambda v : \lambda\in \mathbb{R}_+ \} \subseteq \mathcal C$ is called an \emph{extremal ray} of $\mathcal C$ (we write $\mathbb R_+ v \in \operatorname{ext} \mathcal C$) if
    $$v = x+y \quad\text{with } x,y \in \mathcal C \implies x,y \in \mathbb R_+ v.$$
    \end{definition}
    \begin{definition}
        Given a cone $\mathcal C$, we define its \emph{dual cone} by
    $$\mathcal C^* \:= \{\alpha \in V^* \, : \,\alpha (x) \geq 0, \, \forall x \in \mathcal C\} \subseteq V^*,$$
    where $V^*$ is the vector space dual to $V$.
    If $V$ has an inner product structure, then the elements of the dual are of the form $\alpha (x) = \langle v_\alpha, x \rangle$ for some $v_\alpha \in V$. 
    \end{definition}

\begin{definition}
For convex cones $C_1 \subseteq V$ and $C_2 \subseteq V$ in the real vector space $V$, we define have the sum of cones, 
\[ C_1 + C_2 := \{x+y : x \in C_1, y \in C_2\} \] This is again a convex cone.
\end{definition}

The next theorem is a well known theorem about dual cones and sum of cones in convex geometry. We give the proof for completeness. 
\begin{theorem} \label{thm:dual-cone-sum}
For any convex cone $C_1$ and $C_2$, we have, 
\[ (C_1 + C_2)^\circ  = C_1^\circ \cap C_2^\circ \]
\end{theorem}

\begin{proof}
    Let \( x \in (C_1 + C_2)^\circ \). Then, by definition, \( x \in (C_1 + C_2)^\circ \) implies that \( x \cdot (y_1 + y_2) \geq 0 \) for all \( y_1 \in C_1 \) and \( y_2 \in C_2 \). Setting \( y_2 = 0 \) gives \( x \cdot y_1 \geq 0 \) for all \( y_1 \in C_1 \). Similarly, setting \( y_1 = 0 \) gives \( x \cdot y_2 \geq 0 \) for all \( y_2 \in C_2 \). Therefore, \( x \in C_1^\circ \cap C_2^\circ \).

    To show the converse, assume \( x \in C_1^\circ \cap C_2^\circ \). Then \( x \cdot y_1 \geq 0 \) for all \( y_1 \in C_1 \) and \( x \cdot y_2 \geq 0 \) for all \( y_2 \in C_2 \). For any \( y_1 \in C_1 \) and \( y_2 \in C_2 \), we have \( x \cdot (y_1 + y_2) = x \cdot y_1 + x \cdot y_2 \geq 0 \). Thus, \( x \in (C_1 + C_2)^\circ \). 
\end{proof}

\begin{example}
The cone of \emph{entrywise non-negative} $d\times d$ matrices is defined as
\[ \EWP{d} := \{ A \in \mathcal M_d(\mathbb R) \, :  A_{ij} \geq 0 \,\, \forall i,j \in [d] \}, \]
and the cone of \emph{positive semidefinite} $d\times d$ matrices is defined as
\[\PSD{d} := \{B \in \Msa{d} \, : \, \langle x| B |x \rangle \geq 0 \,\, \forall x \in \mathbb C^d\}\]
play a fundamental role in this work. The two cones $\EWP{d} \subseteq \mathcal M_d(\mathbb R)$ and $\PSD{d} \subseteq \mathcal M_d^{sa}(\mathbb C)$ are self-dual. Their extremal rays are 
\[\operatorname{ext} \EWP{d} = \{\mathbb R_+ \ketbra{i}{j}\}_{i,j\in [d]} \quad \text{ and } \quad \operatorname{ext} \PSD{d} = \{\mathbb R_+ \ketbra{x}{x}\}_{\ket{x} \in \mathbb C^d, \, \ket{x} \neq 0}. \]
\end{example}

\begin{tcolorbox}[
    colback=blue!3,
    colframe=black!50,
    boxrule=0.5mm,
    rounded corners,
    width=\textwidth,
]
In this article, we will deal with some important cones in $\M{d}$ as their sections in the circulant subspace, $\mathsf{Circ}_d \subseteq \M{d}$. Therefore we use the notation 
$$\overrightarrow{\mathcal{K}} = \circulinv{\mathcal{K} \, \cap \, \mathsf{Circ}_d},$$ 
where the cone $\mathcal{K}$ is a cone in $\M{d}$. From the last section we also know that $\mathsf{Circ}_d$ is isomorphic to the space $\C{d}$ by the map $\circul{d} : \C{d} \ni a \mapsto  \circul{a} \in \mathsf{Circ}_d$. 
\end{tcolorbox}
\begin{definition}
Consider the following two cones 
\begin{align*}
    \circEWP{d} &:= \circulinv{\mathsf{Circ}_d \, \cap \, \EWP{d}}\\
    \circPSD{d} &:= \circulinv{\mathsf{Circ}_d \, \cap \, \PSD{d}}
\end{align*}
seen as subsets of $\C{d}$. 
\end{definition}

\begin{proposition}
For the cones $\circEWP{d}$ and $\circPSD{d}$ the following are true. 

    \begin{itemize}
    \item $\operatorname{ext} (\circEWP{d})  = \{ \mathbb{R}^+ \ket{k} \}^{d-1}_{k=0}$ 
    \item $ \operatorname{ext} (\circPSD{d}) = \{ \mathbb{R}^+ \mathcal{F}^{-1}\ket{k}\}^{d-1}_{k=0}$ 
    \end{itemize}
\end{proposition}

\begin{proof}
    The first statement is obvious. For the second part, we observe that $\circPSD{d} = \mathcal{F}^{-1}(\circEWP{d})$, so that $\ext(\circPSD{d}) = \{\mathbb{R}^+\mathcal{F}^{-1}\ket{k}\}^{d-1}_{k=0}$.
\end{proof}

\begin{remark}\label{remark:extcircPSD}
The extremal rays of the $\circPSD{d}$ cone $\mathbb{R}_+ f_k = \mathbb{R}_+ \mathcal{F}^{-1}\ket{k}$, for $k=0,1,\ldots ,d-1$, are just multiples of the columns of the inverse fourier matrix:
\[
(f_k)_i = \frac{1}{\sqrt{d}} \omega^{-ik} = \frac{1}{\sqrt{d}}(f_k * (\overbar{f_k})^{\rev})_i 
\]
\end{remark}

\section{Families of symmetric states}\label{sec:families-symmetric-states}

In this section, we explore different classes of symmetric states that have been previously considered in the literature. Our classification starts from the symmetry group that leaves invariant the bipartite quantum states. We consider unitary representations of a group $G$
$$G \ni g \mapsto U_g \in \mathcal U(d)$$
and the corresponding families of bipartite quantum states $\rho \in \M{d} \otimes \M{d}$ that are invariant under the following actions: 
$$\rho = (U_g \otimes U_g) \rho (U_g \otimes U_g)^* \quad \text{ or } \quad \rho = (U_g \otimes \bar U_g) \rho (U_g \otimes \bar U_g)^*.$$

\subsection{Unitary and orthogonal invariance}

The unitary group $G = \mathcal U(d)$ with its standard representation gives rise to the \emph{isotropic states}
\begin{equation}\label{eq:isotropic-states}
    \rho \in \operatorname{span}\{I_d \otimes I_d, \omega_d\},
\end{equation}
where $\omega_d$ is the \emph{maximally entangled state}
$$\omega_d := \frac 1 d \sum_{i,j=1}^d \ketbra{ii}{jj} \in \M{d} \otimes \M{d},$$
in the case of the $U-U$ representation \cite{Horodecki1999iso}. For the conjugate representation $U-\bar U$, one obtains the \emph{Werner states} \cite{Werner1989}
\begin{equation}\label{eq:Werner-states}
    \rho \in \operatorname{span}\{P_s, P_a\}=\operatorname{span}\{I_d \otimes I_d, F_d\},
\end{equation}
where $P_{s,a}$ are, respectively, the orthogonal projections on the symmetric and anti-symmetric subspaces of $\C{d} \otimes \C{d}$: 
$$P_s = \frac{I_{d^2}+F_d}{2} \qquad P_a = \frac{I_{d^2}-F_d}{2},$$
with $F_d \in \mathcal U(d^2)$ being the \emph{flip} (or \emph{swap}) operator:
$$F_d = \sum_{i,j=1}^d \ketbra{ji}{ij}.$$

For the orthogonal group $G=\mathcal O(d)$, the normal and conjugate representations are identical and give rise to the \emph{Brauer states}
\cite{vollbrecht2001entanglement,keyl2002fundamentals}, which generalize isotropic and Werner states:
\begin{equation}\label{eq:Od-states}
    \rho \in \operatorname{span}\{I_d \otimes I_d, \omega_d, F_d\},
\end{equation}

The separable (or equivalently the $\PPTcone$) states in these classes have been completely characterized \cite{Werner1989} \cite{Horodecki1999iso} \cite{keyl2002fundamentals} ; there are no $\PPTcone$ entangled states.

\subsection{Local diagonal orthogonal invariance}\label{sec:LDOI}

Consider now the subgroup $\mathcal{DU}(d) \leq \mathcal U(d)$ of \emph{diagonal unitary matrices} and its orthogonal counterpart $\mathcal{DO}(d) \leq \mathcal O(d)$. The corresponding invariant states are called respectively local diagonal unitary invariant (LDUI), conjugate local diagonal unitary invariant (CLDUI), and local diagonal orthogonal invariant (LDOI) \cite{singh2021diagonal}. Since the invariance group is smaller that the full unitary (resp.~orthogonal) group, these families are larger then the ones in Eqs.~\eqref{eq:isotropic-states}, \eqref{eq:Werner-states}, and \eqref{eq:Od-states}.

\medskip

\begin{center}
\centering
{\renewcommand{\arraystretch}{1.5}
\begin{tabular}{|r|l|c|} 
\hline
\rowcolor[HTML]{C0C0C0} Acronym    & Symmetry                                   & Condition                                        \\ 
\hline
LDUI & local diagonal unitary invariant           & $(U \otimes U) X (U \otimes U)^* = X$  \\ 
\hline
CLDUI & conjugate local diagonal unitary invariant & $(U \otimes \bar U) X (U \otimes \bar U)^* = X$  \\ 
\hline
LDOI  & local diagonal orthogonal invariant        & $(O \otimes O) X (O \otimes O)^* = X$    \\
\hline
\end{tabular}
}
\end{center}

\medskip

The conditions above hold for all diagonal $d\times d$ unitary matrices $U \in \mathcal{DU}_d$ and all diagonal orthogonal $d\times d$ matrices $O \in \mathcal{DO}_d$. Any LDOI matrix is of the form \cite{singh2021diagonal}
 \begin{align}
    X_{(A,B,C)}  = \sum_{i,j} A_{ij} |ij\rangle\langle ij| + \sum_{i\neq j} B_{ij} |ii\rangle\langle  jj| + \sum_{i\neq j} C_{ij} |ij \rangle\langle  ji|, \label{eq:ABC_LDOI}
 \end{align}
where $A,B,C\in\M{d}$, and $\operatorname{diag}A=\operatorname{diag}B=\operatorname{diag}C$. If $B$ (resp. $C$) here is diagonal, then the resulting family of matrices form the LDUI (resp. CLDUI) subspace. These two subspaces are linked via the operation of partial transpose, and hence the separability results for one class apply identically to the other class as well. CLDUI matrices are of the form:
\begin{align}
     X_{(A,B)}  = \sum_{i,j} A_{ij} |ij\rangle\langle ij| + \sum_{i\neq j} B_{ij} |ii\rangle\langle  jj| \label{eq:AB_CLDUI}
\end{align}

It turns out that separability of these matrices is closely linked with the cones of pairwise and triplewise completely positive matrices, which we introduce now. We denote the entrywise (or Hadamard) product between vectors $v,w\in \C{d}$ by $v\odot w$ or $\ket{v\odot w}$.

\begin{definition} \label{def:PCP-TCP} \cite{singh2021diagonal}
Let $A,B,C\in \M{d}$. 
\begin{itemize}
    \item The pair $(A,B)$ is called \emph{pairwise completely positive} (PCP) if there exist a finite set of vectors $\{\ket{v_k},\ket{w_k}\}_{k\in I}\subset \C{d}$ such that
\begin{align*}
      A = \sum_{k\in I} |v_k \odot \overbar{v_k}\rangle&\langle w_k \odot \overbar{w_k}|, \quad B = \sum_{k\in I} \ketbra{v_k \odot w_k}{v_k \odot w_k} .
\end{align*}
    \item The triple $(A,B,C)$ is called \emph{triplewise completely positive} (TCP) if there exist a finite set of vectors $\{\ket{v_k},\ket{w_k}\}_{k\in I}\subset \C{d}$ such that
\begin{align*}
      A = \sum_{k\in I} |v_k \odot \overbar{v_k}\rangle\langle w_k \odot \overbar{w_k}|, \quad &B = \sum_{k\in I} \ketbra{v_k \odot w_k}{v_k \odot w_k}, \\ &C = \sum_{k\in I} \ketbra{v_k \odot \overbar{w_k}}{v_k \odot \overbar{w_k}}. 
\end{align*}
\end{itemize}

\end{definition}

\begin{theorem} \cite{singh2021diagonal}
Let $A,B,C\in \M{d}$. Then, 
\begin{itemize}
    \item $X_{(A,B)}$ is separable $\iff (A,B)$ is PCP.
    \item $X_{(A,B,C)}$ is separable $\iff (A,B,C)$ is TCP.
\end{itemize}
\end{theorem}

The set of all $d\times d$ PCP and TCP matrix pairs and triples form convex cones \cite[Proposition 5.6]{singh2021diagonal}, which we denote by $\mathsf{PCP}_d$ and $\mathsf{TCP}_d$, respectively.

\subsection{Semi-direct product constructions}

We now consider \emph{intermediate} subgroups
$$\mathcal{DU}(d) \leq G \leq \mathcal U(d)$$
that would give rise to intermediate families of invariant states. To do so, we shall consider semi-direct products of the diagonal unitary group $\mathcal{DU}(d)$ (resp.~ the diagonal unitary group $\mathcal{DU}(d)$) with a subgroup $H$ of the symmetric (permutation) group $\operatorname{Sym}(d)$:
\begin{equation}\label{eq:semi-direct-product}
    G:= H \ltimes \mathcal{DU}(d) \quad \text{ with } H \leq \operatorname{Sym}(d).
\end{equation}
Recall that the semi-direct product $H \ltimes \mathcal{DU}(d)$ group endows the cartesian product $H \times \mathcal{DU}(d) = H \times \mathbb{T}^d$ with the product rule
$$(\sigma, u) \cdot (\pi, v) := (\sigma \pi, u \odot (\sigma.v)),$$
where the action of symmetric group on vectors reads
$$\forall i \in [d]: \qquad (\sigma.v)_i = v_{\sigma^{-1}(i)}.$$
The unitary representation $U_g$ of $G$ is given by permutation matrices with phases. Concretely, for $\sigma\in H$ and $u\in \mathbb{T}^d$,  we have
$$\big[U_{\sigma, u}\big]_{ij} = \mathds{1}_{i = \sigma(j)} u_i.$$
The same notions can be defined for the diagonal orthogonal group $\mathcal{DO}(d)$. Importantly, one recovers the \emph{hyperoctahedral group} as $\operatorname{Hyp}(d):=\operatorname{Sym}(d) \ltimes \mathcal{DO}(d)$. 

The table below shows the states invariant under the action of classical groups that are relevant to our study, using the construction above. For the permutation group $H$, we consider either the full group $\operatorname{Sym}(d)$ or the (abelian) subgroup of \emph{cyclic permutations} $\operatorname{Cyc}(d)$.

\bigskip

\begin{center}
    \bgroup
    \def\arraystretch{1.5}
    \begin{tabular}{|c|c|c|}
        \hline
        \rowcolor[HTML]{C0C0C0} 
        $\ltimes$ & \(\operatorname{Sym}(d) \) & \(\operatorname{Cyc}(d) \) \\
        \hline
        \(\mathcal{D}U(d)\) & axisymmetric states \cite{PhysRevA.94.020302} & generalized axisymmetric states \cite{benkner2022characterizing} \\
        \hline
        \(\mathcal{D}O(d)\) & hyperoctahedral states \cite{park2024universal} & LCSI states (\emph{this paper}) \\
        \hline
    \end{tabular}
    \egroup
\end{center}

\bigskip

 Let us consider in more detail the hyperoctahedral group which can be defined as a semi-direct product of group of permutation matrices with diagonal orthogonal group, $\operatorname{Hyp}(d) = \operatorname{Sym(d)} \ltimes D \mathcal{O}(d)$. A bipartite matrix $X$ is said to be hyperoctahedral if it satisfies $(O\otimes O)X(O \otimes O)^\top = X$ for all orthogonal matrices in the group $\operatorname{Hyp}(d)$. This class of highly symmetric states were considered in the recent paper \cite{park2024universal} where they were shown to have the form 

\begin{equation*}
 X^{\mathsf{Hyp}}_{a,a',b,c} = a \sum_i \ketbra{ii}{ii}  + a'  \sum_{i \neq j} \ketbra{ij}{ij} + b\sum_{i\neq j} \ketbra{ii}{jj} + c \sum_{i\neq j}\ketbra{ij}{ji},
\end{equation*}
for complex parameters $a,a',b,c \in \mathbb C$. This class reduces to the well-known Werner states when $a = a'$ and $b = 0$ (resp.~isotropic states when $a = a'$ and $c = 0$).

In \cite{park2024universal} it was also shown that all the $\PPTcone$ states in this class are also separable.

\begin{theorem}[{{\cite[Theorem 4.1]{park2024universal}}}]
A hyperoctahedral quantum state that is PPT is necessarily separable: 
    $$X^{\mathsf{Hyp}}_{a,a',b,c} \text{   is   }   \SEPcone \iff X^{\mathsf{Hyp}}_{a,a',b,c} \text{   is  } \PPTcone$$ 
\end{theorem}

\begin{remark}
    Instead of taking the complete group of permutation matrices, we can restrict to subgroups that are $2$-transitive (i.e if the orbit of $(\pi(i),\pi(j)) = \{(l,m), l \neq m\ \forall l,m\}$), we get the same class of states.  
\end{remark}

Given the lack of PPT entanglement in the set of hyperoctahedral quantum states, one is naturally led to consider larger families of symmetric states by reducing the size of the symmetry group. One choice is to consider the group cyclic permutations for $H$ in \cref{eq:semi-direct-product}. By replacing the diagonal orthogonal group with the larger group of all diagonal unitary matrices, we obtain the states that are invariant under the group $\CircPhase{d}$, that can be defined as \[\CircPhase{d} := \operatorname{Cyc}(d) \ltimes \mathcal{DU}(d) \]
where $\mathcal{DU}(d)$ is the group of $d \times d$ diagonal unitaries. 

We call these states as Local Cyclic Phase Invariant states (LCPI). These states have also been introduced and studied as a generalization of \emph{axisymmetric states} \cite{PhysRevA.94.020302} in the recent paper \cite{benkner2022characterizing}. They are special cases of LCSI states that we introduce in the next section. 
Importantly, this class of states contain $\PPTcone$ entangled states in all $d \geq 3$ (see \cref{ex:PPT-entangled-LCSI-3} and \cref{rem:LCSI-PPT-entangled}).

\section{Local Cyclic Sign Permutation Invariance}\label{sec:LCSI} 

In this section, we will look at the basic definitions and properties of a new class of bipartite invariant quantum states, which we call \emph{Local Cyclic Sign Invariant} (LCSI) states. The central group in this paper is the group of \textit{cyclic sign permutations}, which we define below and denote by $\CircSign{d}$. We will denote the group of diagonal orthogonal matrices by $\mathcal{DO}_{d}$. This is just a matrix group that includes all orthogonal matrices $O$ such that $O_{ii} = \pm 1$, and $O_{ij} = 0$ for all $i \neq j$.  The group of $d\times d$ cyclic permutation matrices is denoted by $\operatorname{Cyc}_d$. If we define $P = \circul{\ket{1}}\in \M{d}$, then $\operatorname{Cyc}_d$ is the (abelian) group generated by $P$.
\begin{definition}
    The cyclic sign group is defined as 
    $$\CircSign{d}:= \operatorname{Cyc}_d \ltimes \mathcal{DO}_{d} := \{P \cdot O : P \in \operatorname{Cyc}(d) \ \text{and} \ O \in \mathcal{DO}_{d}\}$$
\end{definition}

\begin{tcolorbox}[
    colback=blue!3,
    colframe=black!50,
    boxrule=0.5mm,
    rounded corners,
    width=\textwidth,
]
We can understand the group $\CircSign{d}$  as the group of cyclic permutation matrices, but with the $1$ entries replaced by $\pm 1$. For example, the permutation matrix $P^{d-1}$ gives us

\begin{center}
\begin{tikzpicture}
   \matrix (M1) at (0,0) [matrix of math nodes, left delimiter={[}, right delimiter={]}] {
    0 & 0 & \cdots & 0 & 1 \\
    1 & 0 & \cdots & 0 & 0 \\
    0 & 1 & \cdots & 0 & 0 \\
    \vdots & \vdots & \ddots & \vdots & \vdots \\
    0 & 0 & \cdots & 1 & 0 \\
  };

  \matrix (M2) at (5,0) [matrix of math nodes, left delimiter={[}, right delimiter={]}] {
    0 & 0 & \cdots & 0 & \pm 1 \\
    \pm 1 & 0 & \cdots & 0 & 0 \\
    0 & \pm 1 & \cdots & 0 & 0 \\
    \vdots & \vdots & \ddots & \vdots & \vdots \\
    0 & 0 & \cdots & \pm 1 & 0 \\
  };

  \draw[->, bend right=45] (M1.east) to node[above] {} (M2.west);
\end{tikzpicture}
\end{center}
\end{tcolorbox}

More precisely, recall from the previous section that a general element of the group $\CircSign{d}$ corresponding to a cyclic permutation $\sigma = (1\, 2\, \cdots \, d)^k \in \operatorname{Cyc}(d)$ and a sign vector $\varepsilon \in \{\pm 1\}^d$ is represented by a matrix $U_{\sigma, \varepsilon}$ having elements
$$[U_{\sigma, \varepsilon}]_{ij} = \mathds 1_{i = \sigma(j)} \varepsilon_i = 1_{i = j+k} \varepsilon_i.$$

\begin{remark}
    The group $\CircSign{d}$ is \emph{not} abelian even though both the groups of $\mathrm{Cyc}(d)$ and $D \mathcal{O}(d)$ are abelian:
    $$
    \begin{bmatrix}
        0 & 0 & 1 \\
        1 & 0 & 0 \\
        0 & 1 & 0 
    \end{bmatrix} \cdot
    \begin{bmatrix}
        0 & 0 & -1 \\
        1 & 0 & 0 \\
        0 & 1 & 0 
    \end{bmatrix} =
    \begin{bmatrix}
        0 & 1 & 0 \\
        0 & 0 & -1 \\
        1 & 0 & 0 
    \end{bmatrix} \neq
    \begin{bmatrix}
        0 & -1 & 0 \\
        0 & 0 & 1 \\
        1 & 0 & 0 
    \end{bmatrix} =
    \begin{bmatrix}
        0 & 0 & -1 \\
        1 & 0 & 0 \\
        0 & 1 & 0 
    \end{bmatrix} \cdot
    \begin{bmatrix}
        0 & 0 & 1 \\
        1 & 0 & 0 \\
        0 & 1 & 0 
    \end{bmatrix}.
    $$

\end{remark}

We come now to the main definition of this paper. 

\begin{definition}
\label{lcsi-states}
A bipartite matrix $X \in \M{d} \otimes \M{d}$ is called \emph{Local Cyclic Sign Invariant (LCSI)} if 
\begin{equation*}
 \forall O\in \CircSign{d}: \qquad   (O \otimes O) X (O \otimes O)^\top = X .
\end{equation*}
\end{definition} In the remainder of this section, we will investigate several properties of LCSI matrices. 

\subsection{Linear Structure of LCSI matrices}
\label{sec:linear-space}
Since the condition in Definition~\ref{lcsi-states} is linear, it is easy to see that the set of LCSI matrices (not necessarily quantum states) form a vector subspace of $\M{d} \otimes \M{d}$. In this section, we explicitly characterize the structure of this space in terms of vector triples. Let us begin with a result from \cite{singh2021diagonal} to prove the following lemma.

\begin{lemma}
\label{ldoi-lcsi}
The linear space of $\LCSI{d}$ matrices can be parametrized by $(A,B,C) \in \MLDOI{d}$ such that $A,B,C$ are invariant under the action of $\operatorname{Cyc}(d)$: $PXP^{-1} = X$ for X = $A,B, C$.  
\end{lemma}

\begin{proof}
The linear space of $\LDOI{d}$ matrices can be parameterized using $(A,B,C) \in \MLDOI{d}$ such that any state can be written as 
$$X_{(A,B,C)} = \sum_{ij} A_{ij}\ketbra{ij}{ij} + \sum_{i \neq j} B_{ij}\ketbra{ii}{jj} + \sum_{i \neq j} C_{ij}\ketbra{ij}{ji}$$ 
such that $\diag(A) = \diag(B) = \diag(C)$

We begin the proof by noting that $\LCSI{d} \subseteq \LDOI{d}$. Therefore it is of the form $X_{(A,B,C)}$ and satisfies the invariance condition  
\begin{equation*}
(P \otimes P) X_{A,B,C} (P^{-1} \otimes P^{-1}) =  X_{A,B,C} 
\end{equation*}
Looking at both the sides of the equation, 
\begin{align*}     
     \operatorname{LHS} = &\sum_{ij} A_{ij} \ketbra{\pi(i) \pi(j)}{\pi(i) \pi(j)} + \sum_{ij} B_{ij} \ketbra{\pi(i)\pi(i)}{\pi(j)\pi(j)} 
    + \sum_{ij} C_{ij} \ketbra{\pi(i)\pi(j)}{\pi(j)\pi(i)} \\
    = &\sum_{ij} A_{\pi^{-1}(i)\pi^{-1}(j)}\ketbra{ij}{ij} + \sum_{ij} B_{\pi^{-1}(i)\pi^{-1}(j)}\ketbra{ii}{jj} + \sum_{ij} C_{\pi^{-1}(i)\pi^{-1}(j)}\ketbra{ij}{ji} \\
    \operatorname{RHS} = &\sum_{ij} A_{ij}\ketbra{ij}{ij} + B_{ij}\ketbra{ii}{jj} + C_{ij}\ketbra{ij}{ji}
\end{align*}
This implies that $A_{ij} = A_{\pi^{-1}(i)\pi^{-1}(j)}$ for all $\pi$ in $\operatorname{Cyc}(d)$. This condition can be written as $PAP^{-1} = A$ for all $P$ in $\operatorname{Cyc}(d)$,  and similarly for $B$ and $C$. 
\end{proof}

\begin{proposition}
\label{perm-circulant}
    The set of matrices $X \in \M{d}$ satisfying $PXP^{-1} = X$ for all $P \in \operatorname{Cyc}(d)$ is precisely the set of circulant matrices, i.e. $X_{ij} = a_{i-j}$ for a vector $a \in \C{d}$ having entries $(a_0,a_1, \ldots,  a_{d-1})$ .
\end{proposition}

\begin{proposition}\label{prop:abc}
The linear space of $\LCSI{d}$ is isomorphic to the vector space 
$$(\C{d})^{\times 3}_{\mathbb{C}} := \{(a,b,c) \in (\C{d})^{\times 3} \text{   st   } a_0 = b_0 = c_0\},$$ where the isomorphism can be written as 
$$X_{(a,b,c)}  = \sum_{j,k}a_{k} | {j \oplus k, j} \rangle \langle {j \oplus k, j} | +\sum_{j,k \geq 1} b_{k} | {j \oplus k, j \oplus k} \rangle \langle {j, j} | + \sum_{j,k \geq 1} c_{k} | {j \oplus k, j} \rangle \langle {j, j \oplus k} |$$ 
and $\oplus$ should be understood as sum $\operatorname{mod}(d)$
\end{proposition}

\begin{proof}
    The proof follows directly from Lemma~\ref{ldoi-lcsi} and Proposition~\ref{perm-circulant}. 
\end{proof}

\begin{remark}
The dimension of the complex vector space of $\LCSI{d}$ is $\operatorname{dim}_{\mathbb{C}}(\LCSI{d})= 3d - 2$. The stated isomorphism with vector triples shows that we have $3d$ parameters for $3$ vectors, but since $a_0 = b_0 = c_0$, we get $\operatorname{dim}_{\mathbb{C}}(\LCSI{d})= 3d - 2$. This should be compared with the dimension of $\LDOI{d}$, which scales quadratically in $d$: $\operatorname{dim}_\mathbb{C}(\LDOIcone_d) = 3d^2-2$.
\end{remark}

\subsection{Convex structure of LCSI matrices}
\label{sec:convex-lcsi}
In this subsection, we will look at the cones of positive semi-definite LCSI matrices (by which we will also mean states), the $\PPTcone$ cone and finally the cone of separable matrices. Here, we prove some structure theorems on the vector triple $(a,b,c)$ such that $X_{a,b,c}$ belongs to each of these cones. We look at these cones as they are the most important from the perspective of quantum information theory, particularly in entanglement theory. In the next few sections, we will be interested in exploring 
$\PPTcone$ entangled states and the separability problem in this class of quantum states. The next proposition characterizes membership in the positive semidefinite cone for a matrix and positivity under partial transpose.

\begin{theorem}
\label{theorem:circPPTPSD}
We define, for $k \in [d]$,
\[
\lambda^{\pm}_k := \frac{(a_k + a_{d-k}) \pm \sqrt{(a_k - a_{d-k})^2 + 4 c_{d-k} c_k }}{2}.
\]
Then, the spectrum of $X_{a,b,c}$ is given by:

\[
\operatorname{spec}[X_{a,b,c}] = \left(\sqrt{d}\mathcal{F}b \right) \cup \bigcup_{k=1}^{d-1} \{\lambda^{\pm}_k\}^{\times (d-k)}.
\]

Moreover, for any $X_{a,b,c}$ it holds that: 
\begin{itemize}
    \item $X_{a,b,c} \in \PSD{d^2} \iff a \in \mathbb{R}^{d}_{+}$, $\mathcal{F}b \in \mathbb{R}^{d}_{+}$, $c=\bar c^\rev$ and, $\forall i \in [d]$,  $a_i a_{d-i} \geq \abs{c_i}^2$;
    \item $X_{a,b,c}^\Gamma \in \PSD{d^2} \iff a \in \mathbb{R}^{d}_{+}$, $\mathcal{F}c \in \mathbb{R}^{d}_{+}$, $b=\bar b^\rev$ and, $\forall i \in [d]$, $a_i a_{d-i} \geq \abs{b_i}^2$;
    \item $X_{a,b,c} \in \PPTcone_{d^2} \iff a \in \mathbb{R}^{d}_{+}$, $\mathcal{F}b, \mathcal{F}c \in \mathbb{R}^{d}_{+}$ and, $\forall i \in [d]$, $a_i a_{d-i} \geq \operatorname{max} (\abs{b_i}^2,\abs{c_i}^2)$.
\end{itemize}
\end{theorem}

\begin{proof}

We can use the following block decomposition of the matrix $X_{a,b,c}$ to prove these results, see \cite[Proposition 4.1]{singh2021diagonal}:
\[
X_{a,b,c} = \circul{b} \oplus {\bigoplus^{d-1}_{k=1}} \begin{pmatrix}
a_k & c_k \\
c_{d-k} & a_{d-k}
\end{pmatrix}^{\oplus (d-k)}.
\]  

For any circulant matrix $\circul{b} \in \M{d}$, we have $\operatorname{spec}[\circul{b}] = \sqrt{d}\mathcal{F} (b)$. Finally, from \cite{singh2021diagonal}, we know that $X_{a,b,c}^{\Gamma} = X_{a,c,b}$.
\end{proof} 
Next, we look at the separable cone ($\SEPcone$). Recall that ${(a * b)}_k  = \sum_i a_i b_{k - i}$. For $a\in \C{d}$, we define the reflected vector $a^\rev \in \C{d}$ as ${a}^\rev_i = a_{d-i}$. The map $a \rightarrow \overbar{a} * a^\rev$ is called the autocorrelation map. We now derive the necessary and sufficient conditions for the separability of the $\LCSIcone$ matrix $X_{a,b,c}$. For this we introduce another notion of positivity for vectors $a,b,c$ which will be called \emph{Circulant} Triplewise Completely Positive, inspired by the cone of Triplewise Completely Positive matrices introduced in \cite[Definition 7.4]{nechita2021graphical}, which, in turn, generalizes \cite[Definition 3.1]{johnston2019pairwise}. 

\begin{definition}\label{def:circTCP}
  \label{TCP}
    A vector triple $(a,b,c)\in (\C{d})^{\times 3}_\mathbb{C}$ is called \emph{Circulant Triplewise Completely Positive} if there exist a finite set of vectors $\{v_k, w_k\}_{k\in I}$ such that 
    
\begin{align*}
a = \sum_{k\in I} ({v_k \odot \overbar{v_k})} * (\overbar{w_k} \odot w_k)^\rev, \qquad
b &= \sum_{k\in I} {(v_k \odot w_k)} * ( \overbar{v_k} \odot \overbar{w_k})^\rev, \\
c &= \sum_{k\in I} {(v_k \odot \overbar{w_k})} * (\overbar{v_k} \odot w_k)^\rev,
\end{align*}
where we recall that $*$ denotes the circular convolution and $\odot$ denotes the Hadamard (entrywise) product of vectors. We denote the set of all such vector triples $(a,b,c)$ by $\circTCP{d}$.
\end{definition}

\begin{theorem}
The following holds true for $\LCSIcone$ matrices:
$$X_{a,b,c} \text{ is separable } \iff (a,b,c) \in {\circTCP{d}}.$$
\end{theorem}
\begin{proof}

Following Section~\ref{def:PCP-TCP}, we know that for a separable LDOI matrix $X^{\mathsf{LDOI}}_{(A,B,C)}$, there exist finite set of vectors $v_k$ and $w_k$ such that 
\begin{align*}
\label{def:TCP}
        A &= \sum_k \ketbra{v_k \odot \overbar{v_k}}{w_k \odot \overbar{w_k}} \quad
        B = \sum_{k} \ketbra{v_k \odot w_k} \quad
        C = \sum_k \ketbra{v_k \odot \overbar{w_k}}
    \end{align*}

From Section~\ref{sec:linear-space}, we know that $\LCSIcone$ states with the triple $a,b,c$ are $\LDOIcone$ states with $A = \circul{a}, B = \circul{b}$ and $C = \circul{c}$. Recall that $\circul{a} = \sum_{i=0}^{d-1} a_i P^i$, where $P$ is the shift permutation from Remark~\ref{Remark:shift} satisfying $P^i P^j = P^{i \oplus j }$ and $\operatorname{Tr}(P^i) = d \mathds{1}_{i = 0 \, \textrm{mod} \, d}$. Hence, we obtain
\begin{align*}
\sum_{i=0}^{d-1} a_i P^i &= \sum_k \ketbra{\overbar{v_k} \odot v_k}{\overbar{w_k} \odot w_k}.
\end{align*}
Taking the trace of both sides after multiplying by \(P^{-j}\), we have:
\begin{align*}
d \sum_{i=0}^{d-1} a_i \delta_{i-j} =\text{Tr}\left(\sum_{i=0}^{d-1} a_i P^i P^{-j}\right) &=  
\sum_{i=0}^{d-1} a_i \text{Tr}(P^i P^{-j}) \\ 
&= \sum_k \text{Tr}\left(\ketbra{\overbar{v_k} \odot v_k}{\overbar{w_k} \odot w_k} P^{-j}\right) \\
&= \sum_k \bra{\overbar{w_k} \odot w_k} P^{-j} \ket{ \overbar{v_k} \odot v_k}.
\end{align*}
This simplifies to
\begin{align*}
a = \frac{1}{d} \sum_k (w_k \odot \overbar{w_k}) * (\overbar{v_k} \odot v_k)^{\rev}.
\end{align*}

We can do a similar calculation for $B = \operatorname{circ}(b)$ and $C = \operatorname{circ}(c)$ to show that the vectors $(a,b,c)$ form a $\circTCP{d}$ triple of the form in Definition~\ref{TCP}.

To show the converse, we begin with $(a,b,c) \in \circTCP{d}$ that is of the form given in Definition~\ref{TCP} and show that $(\circul{a},\circul{b},\circul{c}) \in \operatorname{TCP}_d$. Again, we will do an explicit calculation for $a$, and a similar calculation can be done for $b$ and $c$ to show that $(\circul{a},\circul{b},\circul{c})$ is of the form in Definition~\ref{def:PCP-TCP}. We start with the given expression:
\begin{align*}  
a &= \sum_k ({v_k \odot \overbar{v_k}}) * ({\overbar{w_k} \odot w_k})^{\rev} \implies 
a_j = \sum_k \bra{\overbar{v_k} \odot v_k} P^{-j} \ket{\overbar{w_k} \odot w_k}.
\end{align*}

Now, consider the circulant matrix generated by \(a\):
\begin{align*}
\circul{a} &= \sum_{i=0}^{d-1} a_i P^i \\
&= \sum_k \sum_{i=0}^{d-1} \bra{\overbar{v_k} \odot v_k} P^{-i} \ket{\overbar{w_k} \odot w_k} P^i \\
&= \sum_k \sum_{i=0}^{d-1} \text{Tr} \left(\ketbra{\overbar{w_k} \odot w_k}{\overbar{v_k} \odot v_k} P^{-i}\right) P^i \\
&= \sum_k \sum_{i=0}^{d-1} \frac{1}{d} \text{Tr} \left( \sum_j P^j \ketbra{\overbar{w_k} \odot w_k}{\overbar{v_k} \odot v_k} P^{-j} P^{-i} \right) P^i \\
&=  \sum_k \sum_j P^j \ketbra{\overbar{w_k} \odot w_k}{\overbar{v_k} \odot v_k} P^{-j} \\
&= \sum_{k,j} \ketbra{\overbar{w_{kj}} \odot w_{kj}}{\overbar{v_{kj}} \odot w_{kj}},
\end{align*}
where \(v_{kj} = P^j | v_k \rangle \) and \(w_{kj} = P^j |w_k \rangle \).
\end{proof}

\section{Cyclic mixtures of Dicke states}\label{sec:cyclic-Dicke}

In this section we focus on a special subset of bipartite quantum states with circular symmetry, those corresponding to \emph{mixtures of Dicke states}. Such states have received a lot of attention in the general (no circulant symmetry) case \cite{tura2018separability,yu2016separability,singh2021diagonal}, in particular due to the connection to the theory of \emph{completely positive matrices} \cite{berman2003completely}. 

\begin{definition}
    A state $X \in \M{d} \otimes \M{d}$ is said to be a \emph{mixture of Dicke states} if it can be written as 
    \begin{equation}\label{eq:Dicke-mixture}
        X = \sum_{i,j=1}^d p_{ij} \ketbra{D_{ij}}{D_{ij}}
    \end{equation}
    where $D_{ij} = \frac{1}{\sqrt{2}}(|ij\rangle + |ji\rangle)$ for $i \neq j$, $D_{ii} = |ii\rangle$, $p_{ij} = p_{ji} \geq 0$, and $\sum_{i,j=1}^d p_{ij} = 1$. 
\end{definition}

It has been well understood that characterizing PPT and separability for Dicke states reduce to checking membership in the cones of doubly non-negative and completely positive matrices, respectively \cite{tura2018separability,yu2016separability}. Recall that a matrix $A\in \M{d}$ is called \emph{completely positive} if it admits a decomposition $A=\sum_i \ketbra{v_i}$ such that for each $i$, $v_i \in \R{d}_+$ \cite{berman2003completely}. The cone of all completely positive $d\times d$ matrices is denoted $\CP{d}$. The cone of $d\times d$ \emph{doubly non-negative} matrices is defined as $\DNN{d}:= \EWP{d}\cap \PSD{d}$. Clearly, $\CP{d}\subseteq \DNN{d}$, where equality holds if and only if $d\leq 4$ \cite{berman2003completely}. Note that Dicke states form a subclass of LDOI states with the corresponding matrix triples $(A,B,C)$ satisfying $A_{ij} = C_{ij} = p_{ij}$ and $B = \text{diag}(A)$ \cite{singh2021diagonal}, which can be used to prove the following result. 

\begin{theorem}\cite{tura2018separability}
    The following equivalences hold true for Dicke states:
    \begin{itemize}
        \item $X_{(A,\diag{A},A)}$ is separable $\iff$ $(A, \diag{A}, A)\in \mathsf{TCP}_d \iff A\in \CP{d}$.
        \item $X_{(A,\diag{A},A)}$ is PPT $ \iff A\in \DNN{d}$.
    \end{itemize}
\end{theorem}

Similarly, mixtures of Dicke states with the following additional symmetry,
$$(P \otimes P) X (P \otimes P)^* = (P \otimes P) X (P \otimes P)^\top =  X$$ for  all cyclic permutation matrices $P \in \operatorname{Cyc}(d)$ can be understood as the subclass of $\LCSI{d}$ states (see \cref{sec:LCSI}) with 
$$a_k = c_k = p_{i,i+k}, \qquad  \forall i,k \in [d].$$
In particular, this means that the symmetric matrix $p$ defining the Dicke state mixture in \cref{eq:Dicke-mixture} is circulant, and the vectors $a,c$ in the LCSI writing are given by the first row of $p$, while the $b$ vector is trivial (only its $0$-th coordinate being non-zero). 

Below is the general form of a circulant mixture of Dicke states, in the cases $d=2,3$ (dots represent $0$ entries): 
\begin{center}
    $$
    X_{(a_0,a_1),(a_0,0),(a_0,a_1)} = \left[
    \begin{array}{ *{2}{c} | *{2}{c} }
       a_0 & \cdot & \cdot & \cdot \\
       \cdot & a_1 & a_1 & \cdot  \\ \hline
       \cdot & a_1 & a_1 & \cdot  \\
       \cdot & \cdot & \cdot & a_0
      \end{array}
    \right]
    $$
    $$  
    X_{(a_0,a_1,a_2),(a_0,0,0),(a_0,a_1,a_2)} = \left[
\begin{array}{ *{3}{c} | *{3}{c} | *{3}{c} }
   a_0 & \cdot & \cdot & \cdot & \cdot & \cdot & \cdot & \cdot & \cdot \\
   \cdot & a_1 & \cdot & a_1 & \cdot & \cdot & \cdot & \cdot & \cdot \\
   \cdot & \cdot & a_2 & \cdot & \cdot & \cdot & a_2 & \cdot & \cdot \\\hline
   \cdot & a_2 & \cdot & a_2 & \cdot & \cdot & \cdot & \cdot & \cdot \\
   \cdot & \cdot & \cdot & \cdot & a_0 & \cdot & \cdot & \cdot & \cdot \\
   \cdot & \cdot & \cdot & \cdot & \cdot & a_1 & \cdot & a_1 & \cdot \\ \hline
   \cdot & \cdot & a_1 & \cdot & \cdot & \cdot & a_1 & \cdot & \cdot \\
   \cdot & \cdot & \cdot & \cdot & \cdot & a_2 & \cdot & a_2 & \cdot \\
   \cdot & \cdot & \cdot & \cdot & \cdot & \cdot & \cdot & \cdot & a_0 \\
  \end{array}
\right].
    $$
\end{center}

We define the cones $\circCP{d}$ and $\circDNN{d}$ as the intersection of the cones $\CP{d}$ and $\DNN{d}$ with the circulant subspace:
\begin{align*}
    \circCP{d} := \cone\{v * v^\rev : v \in \mathbb{R}^d_+\} \quad \text{and} \quad
    \circDNN{d} := \cone \{a \in \mathbb{R}_+^d \, : \, \mathcal{F}a \in \mathbb{R}_+^d\} = \mathbb R^d_+ \cap \mathcal F \mathbb R^d_+.
\end{align*}

\begin{proposition}

    For all $d \geq 2$, we have $\circCP{d} \subseteq \circDNN{d} \subseteq \mathbb R^d_+$.
\end{proposition}
\begin{proof}
    Consider an arbitrary element of the cone $\circCP{d}$
    $$a = \sum_k \lambda_k v_k * v_k^\rev$$
    for non-negative vectors $v_k \in \mathbb R^d_+$ and non-negative scalars $\lambda_k \geq 0$. Clearly, $a$ is entrywise non-negative. 
    Taking the Fourier transform of $a$ we have 
    $$ \sqrt{d} \mathcal{F}  a  = \sqrt{d} \sum_k \lambda_k \mathcal{F}(v_k * v_k^\rev) = d\sum_k \lambda_k  \mathcal{F}v_k \odot \mathcal{F}(v_k^R) = d\sum_k  \lambda_k \mathcal{F}v_k \odot \overbar{\mathcal{F} v_k } \in 
    \mathbb{R}_+^d,$$
    proving that $\mathcal F a$ is also entrywise non-negative and finishing the proof. 
 
\end{proof}

The connection to separability and the PPT property of the corresponding LCSI states are given in the following result. 

\begin{proposition}
The following equivalences are true for vector triples of the form $(a,a_0 \ket{0},a)$

\begin{itemize}
    \item $X_{a,a_0 \ket{0},a} \in \SEPcone \iff (a,a_0 \ket{0},a) \in \circTCP{d} \iff a \in \circCP{d}$
    \item $X_{a,a_0 \ket{0},a} \in \PPTcone \iff a \in \circDNN{d}$
\end{itemize}

\end{proposition}

The remainder of this section is devoted to characterizing the geometry of these cones. Importantly, for every $d\geq 5$, there exist $\circCP{d}$ vectors that are not in $\circDNN{d}$, signaling the presence of PPT entanglement in the class of symmetric states we consider.  

\begin{remark}
    The mixture of Dicke states are a class of bosonic mixed states. For this class of states, all the usual entanglement tests coincide, the PPT criterion, the realignment, the covariance matrix criterion coincide. 
\end{remark}

\subsection{PPT and semi-positive cones}

In this section we are considering cones $X \subseteq \R{d}$ such that the $\mathcal F\cdot X \subseteq \R{d}$ or equivalently we are looking at vectors $X \ni a = a^\rev$. These conditions impose linear constraints on the cones, hence they have, in general, empty interior in $\R{d}$. To remedy the situation, we shall consider their linear closure, hence reducing the total dimension of the underlying vector space. 

Let us introduce the new parameter $n := 1+\lf d/2 \rf$, which counts the number of free parameters of reflection-invariant real vectors. We can construct an orthonormal basis $\{f_j\}_{j \in [n]}$for the space 
$$E_d :=\{a \in \R{d} \, : \, a = a^\rev\}$$
as follows:
\begin{equation}\label{eq:basis-change-F-G}
\{e_0\} \cup \Big\{\frac{e_i + e_{-i}}{\sqrt{2}}\Big\}_{i=1}^{n-2} \cup \{e_{n-1} \}\quad \text{if $d$ is even;} \qquad\qquad
\{e_0\} \cup \Big\{\frac{e_i + e_{-i}}{\sqrt{2}}\Big\}_{i=1}^{n-1} \quad \text{if $d$ is odd.}
\end{equation}

The space $E$ is left invariant by the Fourier matrix $\mathcal F_{ij} = \omega^{ij}/\sqrt d$. We denote by $G$ the restriction of $\mathcal F$ to the space $E$. One can compute the matrix elements of $G$ from those of $\mathcal F$. For example, in the case where $f_j = (e_j + e_{-j})/\sqrt 2$: 

$$\bra{f_0}\mathcal F\ket{f_j} = \frac{\bra{e_0}\mathcal F\ket{e_j} + \bra{e_0}\mathcal F\ket{e_{-j}}}{\sqrt{2}} = \sqrt{2} \Re \bra{e_0}\mathcal  F\ket{e_j}.
$$
Clearly, $G$ is a symmetric matrix ($G = G^\top)$ and we have 
\[
G_{ij} =
\begin{cases}
    \frac{1}{\sqrt{d}} & \text{if } i = j = 0, \\[8pt]
    \sqrt{\frac{2}{d}} & \text{if } i = 0 \text{ and } 1 \leq j < \frac{d}{2}, \\[8pt]
    \frac{1}{\sqrt{d}} & \text{if } i = 0 \text{ and } j = \frac{d}{2} \text{ (for even \(d\))}, \\[8pt]
    \frac{2}{\sqrt{d}} \cos\left( \frac{2 \pi i j}{d} \right) & \text{if } 1 \leq i,j < \frac{d}{2}, \\[8pt]
    \sqrt{\frac{2}{d}} (-1)^i & \text{if } 1 \leq i < \frac{d}{2} \text{ and } j = \frac{d}{2} \text{ (for even \(d\))}, \\[8pt]
    \frac{(-1)^{d/2}}{\sqrt{d}} & \text{if } i = j = \frac{d}{2} \text{ (for even \(d\))}.
\end{cases}
\]

For $d$ even, we have
\[
G =
\frac{1}{\sqrt{d}} \begin{pmatrix}
1 & \sqrt{2} & \sqrt{2} & \cdots & \sqrt{2} & 1 \\
\sqrt{2} & 2 \cos\left(\frac{2\pi \cdot 1 \cdot 1}{d}\right) & 2 \cos\left(\frac{2\pi \cdot 1 \cdot 2}{d}\right) & \cdots & 2 \cos\left(\frac{2\pi \cdot 1 \cdot \frac{d}{2}-1}{d}\right) & \sqrt{2} (-1)^1 \\
\sqrt{2} & 2 \cos\left(\frac{2\pi \cdot 2 \cdot 1}{d}\right) & 2 \cos\left(\frac{2\pi \cdot 2 \cdot 2}{d}\right) & \cdots & 2 \cos\left(\frac{2\pi \cdot 2 \cdot \frac{d}{2}-1}{d}\right) & \sqrt{2} (-1)^2 \\
\vdots & \vdots & \vdots & \ddots & \vdots & \vdots \\
\sqrt{2} & 2 \cos\left(\frac{2\pi \cdot (\frac{d}{2}-1) \cdot 1}{d}\right) & 2 \cos\left(\frac{2\pi \cdot (\frac{d}{2}-1) \cdot 2}{d}\right) & \cdots & 2 \cos\left(\frac{2\pi \cdot (\frac{d}{2}-1) \cdot \frac{d}{2}-1}{d}\right) & \sqrt{2} (-1)^{\frac{d}{2}-1} \\
1 & \sqrt{2} (-1)^1 & \sqrt{2} (-1)^2 & \cdots & \sqrt{2} (-1)^{\frac{d}{2}-1} & (-1)^{\frac{d}{2}}
\end{pmatrix}.
\]

For $d$ odd, we have

\[
G =
\frac{1}{\sqrt{d}} \begin{pmatrix}
1 & \sqrt{2} & \sqrt{2} & \cdots & \sqrt{2} \\
\sqrt{2} & 2 \cos\left(\frac{2\pi \cdot 1 \cdot 1}{d}\right) & 2 \cos\left(\frac{2\pi \cdot 1 \cdot 2}{d}\right) & \cdots & 2 \cos\left(\frac{2\pi \cdot 1 \cdot \frac{d-1}{2}}{d}\right) \\
\sqrt{2} & 2 \cos\left(\frac{2\pi \cdot 2 \cdot 1}{d}\right) & 2 \cos\left(\frac{2\pi \cdot 2 \cdot 2}{d}\right) & \cdots & 2 \cos\left(\frac{2\pi \cdot 2 \cdot \frac{d-1}{2}}{d}\right) \\
\vdots & \vdots & \vdots & \ddots & \vdots \\
\sqrt{2} & 2 \cos\left(\frac{2\pi \cdot \frac{d-1}{2} \cdot 1}{d}\right) & 2 \cos\left(\frac{2\pi \cdot \frac{d-1}{2} \cdot 2}{d}\right) & \cdots & 2 \cos\left(\frac{2\pi \cdot \frac{d-1}{2} \cdot \frac{d-1}{2}}{d}\right)
\end{pmatrix}.
\]

Below are the matrices $G$, for $d=2,3,4,5$ (and, respectively, $n=2,2,3,3$):
$$G^{(d=2)} = \frac{1}{\sqrt 2}\begin{bmatrix}
    1 & 1 \\
    1 & -1
\end{bmatrix} \qquad 
G^{(d=3)} = \frac{1}{\sqrt 3}\begin{bmatrix}
    1 & \sqrt 2 \\
    \sqrt 2 & -1
\end{bmatrix}
$$
$$G^{(d=4)} = \frac{1}{2}\begin{bmatrix}
    1 & \sqrt 2 & 1 \\
    \sqrt 2 & 0 & -\sqrt 2 \\
    1 & -\sqrt 2 & 1
\end{bmatrix} \qquad 
G^{(d=5)} = \frac{1}{\sqrt 5}\begin{bmatrix}
    1 & \sqrt 2 & \sqrt 2 \\
    \sqrt 2 & \frac{\sqrt 5-1}{2} & \frac{-\sqrt 5-1}{2} \\
    \sqrt 2 & \frac{-\sqrt 5-1}{2} & \frac{\sqrt 5-1}{2}
\end{bmatrix}.
$$

We shall identify in what follows the vector space $E_d$ with $\R{n}$ (recall that $n = 1 + \lf d/2 \rf$). Therefore, to every $a \in \circDNN{d}$, we have assign  $b \in \mathbb{R}^{n}$:
$$a \leftrightarrow b \iff \mathcal Fa \leftrightarrow Gb$$

In this section, we prove some results to understand the geometry of the convex cones $\circDNN{d}$ and $\circCP{d}$ that characterize completely the cones of $\SEPcone$ and $\PPTcone$ in the class of states that we introduced in the preceding section. We will look at the convex geometry of these cones as subsets of $\mathbb{R}^n$. 

Since $\circDNN{d}$ is a polyhedral cone, it is generated by a finite number of extreme rays. As we shall see, the analytical enumeration of all extreme rays of $\circDNN{d}$ is still difficult problem. Here we prove certain results about this cone, and provide a simple algorithm to calculate its extreme rays. In this section, we shall continue using the correspondence between $\circDNN{d} \subseteq \R{d}$ and the real vector space $E_d \approx \R{n}$. We have 
$$\circDNN{d} \leftrightarrow \{b \in \R{n}_+ \, : \, G b \in \R{n}_+\}.$$

We provide some results about the extremal rays of semi-positive cones in \cref{sec:ext-vecDNN} and use it to study the \emph{extremal symmetric PPT states}.

\subsection{Geometry of PPT entangled states}

This section contains the exact description of the cones $\circCP{d}$ for $d \leq 5$ and some partial results for $d=6,7$. The main result is the complete characterization of the cone $\circCP{5}$ (and also of its dual), that is not equal to the larger cone $\circDNN{5}$, see \cref{fig:circDNN-circCP-d5}, which allows us to construct examples of $\PPTcone$ entangled states. 
\subsubsection{$d\leq 4$}

In this case, the following proposition for the equality of the two sets (separable and $\PPTcone$) can be shown 
\begin{proposition}
\label{prop:dleq4}
For $d \leq 4$, $\circCP{d} =
\circDNN{d}$: a circulant mixture of Dicke states is separable if and only if it is PPT. 
\end{proposition}
\begin{proof}
    This follows from the more general statement in \cite{tura2018separability,yu2016separability} that (general) mixtures of Dicke states in local dimension 4 or less are separable if and only if they are PPT. In turn, this is a consequence of the well-know fact that a matrix of size 4 or less is completely positive if and only if it is positive semidefinite and entrywise positive \cite[Theorem 3.35]{shaked2021copositive}.
\end{proof}

\begin{figure}[h!]
    \centering
    \includegraphics[width=.5\textwidth]{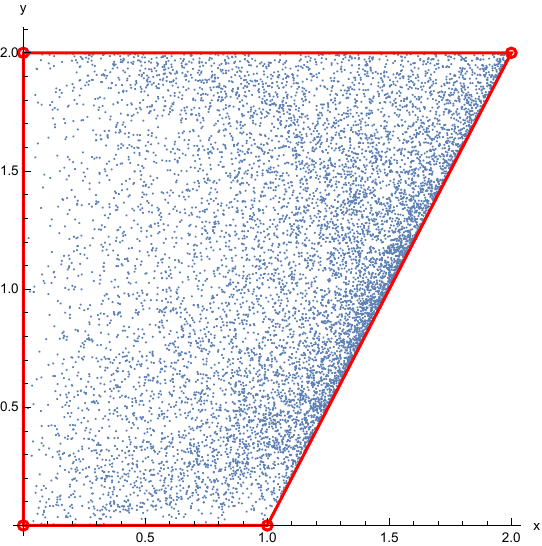}
    \caption{The slice $(2,x,y,x)$ through the cones $\circDNN{4}=\circCP{4}$. In blue, $10^4$ randomly generated points inside $\circCP{4}$; in red, the 4 extreme points of $\circDNN{4}$ from \cref{sec:facets-circDNN} and the polyhedral slice they generate}
    \label{fig:circDNN-circCP-d4}
\end{figure}

We display in \cref{fig:circDNN-circCP-d4} a slice through these cones, showing how randomly generated elements from the $\circCP{4}$ cone slice fill the polyhedron spanned by the 4 extremal elements of the $\circDNN{4}$ cone from \cref{sec:facets-circDNN}.

\subsubsection{$d = 5$}
\begin{figure}[htb]
    \centering
    \includegraphics[width=.5\textwidth]{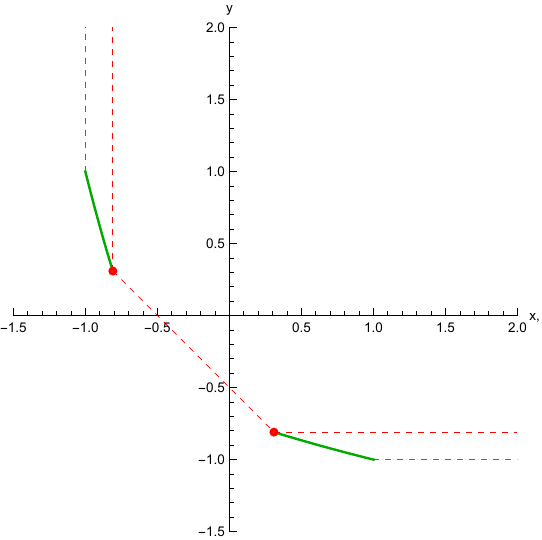}
    \caption{The slice $(1,x,y,y,x)$ through the cones $\circSPN{5} \subsetneq \circCOP{5}$. The red dots are the 2 extreme points $(1,-\cos(\pi/5), \cos(2\pi/5),\cos(2\pi/5),-\cos(\pi/5))$, $(1,\cos(2\pi/5), -\cos(\pi/5),-\cos(\pi/5),\cos(2\pi/5))$ of $\circSPN{5}$ belonging to this slice. The red dashed lines correspond to the boundary of the (unbounded) slice. Solid green curves correspond to the extremal rays of $\circCOP{5}$ belonging to this slice: $h_\theta$ and $h'_\theta$ for $\theta \in [0,\pi/5]$. Dashed green lines depict the boundary of $\circCOP{5}$.}
    \label{fig:circSPN-circCOP-d5}
\end{figure}

\begin{figure}[htb]
    \centering
    \includegraphics[width=.5\textwidth]{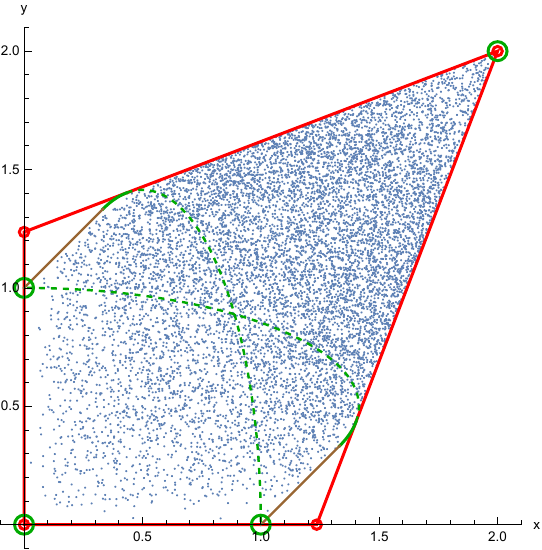}
    \caption{The slice $(2,x,y,y,x)$ through the cones $\circCP{5} \subsetneq \circDNN{5}$. In blue, $10^4$ randomly generated points inside $\circCP{5}$. In red, the 4 extreme points of $\circDNN{5}$ from \cref{sec:facets-circDNN} and the polyhedral slice they generate. Solid green curves correspond to the extremal rays $x_\theta$ and $x'_\theta$ for $\theta \in [0,\pi/5]$, while the three green points are the extremal rays in the directions $(1,0,0,0,0)$, $(1,1,1,1,1)$, $(2,1,0,0,1)$, $(2,0,1,1,0)$. Dashed green lines are non-extremal elements of $\circCP{5}$ corresponding to $x_\theta$ and $x'_\theta$ in the parameter range $\theta \in (\pi/5,\pi/2]$. The two brown lines fill in the missing (non-extremal) part of the boundary of $\circCP{5}$.}
    \label{fig:circDNN-circCP-d5}
\end{figure}

In this section, we discuss the geometry of the set $\circCP{d}$ in the simplest non-trivial case $d=5$. Indeed, for $d \leq 4$, since $\circCP{d} = \circDNN{d}$, the $\circCP{d}$ cone is a polyhedron that was completely described via its (finitely many) extremal rays in the previous two sections. For $d \geq 5$, the $\circCP{d}$ cone has more complex structure inside the $\circDNN{d}$ which is still a polyhedral cone. We shall completely characterize the geometry of the $\circCP{5}$ cone and its dual, providing a list of its (infinitely many) extremal rays. The main results of this section are \cref{thm:circCOP-d5} and \cref{thm:circCP-d5}, which are summarized in \cref{fig:circSPN-circCOP-d5} and \cref{fig:circDNN-circCP-d5} respectively. To discuss the more interesting cases of $d \geq 5$, let us first introduce in the general case the dual objects needed in our analysis.
 
\begin{definition}
    We can define the dual cone of $\circCP{d}$ as follows: 
    $$(\circCP{d})^* = \{a \in \R{d} \, : \,    \langle a, (v * v^\rev) \rangle \geq 0 \quad \forall \,  v \in \mathbb{R}^{d}_+\}.$$ 
\end{definition}

Recall that \emph{copositive matrices} \cite{shaked2021copositive} are the dual of completely positive matrices: 
$$\COP{d} := \{A \in \Msareal{d} \, : \, \langle v, Av \rangle \geq 0 \quad \forall v \in \R{d}_+\}.$$

The following proposition simply states that vectors in the dual of $\circCP{d}$ correspond to circulant copositive matrices; we leave the proof to the reader.
\begin{proposition}
    We have 
    $$(\circCP{d})^* =  \circulinv{\mathsf{Circ}_d  \, \cap \, \COP{d}}=:\circCOP{d}.$$ 
\end{proposition}

Therefore, elements of $\circCOP{d}$ behave like entanglement witnesses for circulant mixtures of Dicke states. An important example of a circulant copositive matrix is the \emph{Horn matrix} \cite{hall1963copositive}: 
\begin{equation}\label{eq:Horn-matrix}
H = \circul{1,-1,1,1,-1} = \begin{pmatrix*}[r]
    1 & -1 & 1 & 1 & -1 \,\,\, \\
    -1 & 1 & -1 & 1 & 1 \,\,\, \\
    1 & -1 & 1 & -1 & 1 \,\,\, \\
    1 & 1 & -1 & 1 & -1 \,\,\, \\
    -1 & 1 & 1 & -1 & 1 \,\,\,
\end{pmatrix*} \in \COP{5}.
\end{equation}
Note that elements in $\circCOP{d}$ can have negative elements; there are however some simple necessary conditions for membership in $\circCOP{d}$ that we gather in the following lemma. 

\begin{lemma}\label{lem:circCOP-necessary-positive}
    Let $a \in \circCOP{d}$. Then 
    \begin{itemize}
        \item $a_0 \geq 0$;
        \item for all $k \in [d]$, $2a_0 + a_{2k}+a_{-2k} \geq 0$.
    \end{itemize}
\end{lemma}
\begin{proof}
    The result follows from the definition of the set $\circCOP{d}$ using vectors $v$ of the respective forms:
    \begin{itemize}
        \item $v=\ket k$
        \item $v = \ket k + \ket{-k}$.
    \end{itemize}
\end{proof}

One can define similarly the dual cone of $\circDNN{d}$:
$$(\circDNN{d})^* =  \circulinv{\mathsf{Circ}_d  \, \cap \, (\PSD{d}^{\mathbb R} + \EWP{d}))}=:\circSPN{d}.$$ 
We call such matrices \emph{circulant $\mathsf{SPN}$} matrices, where 
$$\mathsf{SPN}_d = \PSD{d}^{\mathbb R} + \EWP{d}$$
is the cone of $\mathsf{SPN}$ matrices, see \cite[Theorem 1.167]{shaked2021copositive}.

\begin{proposition}
    The cone $\circSPN{5}$ has 4 extremal rays, generated by the following vectors: 
    \begin{align*}
        &(0,1,0,0,1), (0,0,1,1,0), \\
        &(1,-\cos(\pi/5), \cos(2\pi/5),\cos(2\pi/5),-\cos(\pi/5)),\\
        &(1,\cos(2\pi/5), -\cos(\pi/5),-\cos(\pi/5),\cos(2\pi/5)).
    \end{align*}
\end{proposition}
\begin{proof}
    Recall from the previous sections that the $\circDNN{d}$ cone was defined via the conditions $a \in \R{d}_+$ and $\mathcal F a \in \R{d}_+$. Hence, there are at most $2(1 + \lfloor d/2 \rfloor)=6$ extremal rays of $\circSPN{5}$:
        \begin{align*}
        &(1,0,0,0,0),(0,1,0,0,1), (0,0,1,1,0), (1,1,1,1,1),\\
        &(1,-\cos(\pi/5), \cos(2\pi/5),\cos(2\pi/5),-\cos(\pi/5)),\\
        &(1,\cos(2\pi/5), -\cos(\pi/5),-\cos(\pi/5),\cos(2\pi/5)).
    \end{align*}
    One can easily see that the first and the fourth elements in the list above can be obtained by positive linear combinations of the four others and that the remaining four rays are extreme.
\end{proof}

We characterize \emph{extremal} circulant copositive matrices (or circulant entanglement witnesses) in local dimension $d=5$ in the result below, see also \cref{fig:circSPN-circCOP-d5}.

\begin{theorem}\label{thm:circCP-d5}
Define, for $\theta \in \mathbb R$, the vectors
    \begin{align*}
        x_\theta &:= (2\cos(2\theta)+4, 4\cos \theta, 1, 1, 4 \cos \theta)\\
        x'_\theta &:= (2\cos(2\theta)+4, 1, 4\cos \theta, 4 \cos \theta, 1).
    \end{align*}
The extreme rays of the $\circCP{5}$ cone are given by: 
\begin{align*}
	\operatorname{ext} \circCP{5} = \mathbb R_+ \cdot \big[ &\{(1,0,0,0,0), (1,1,1,1,1), (2,1,0,0,1), (2,0,1,1,0)\} \sqcup \\
	& \qquad \{x_\theta \, : \, \theta \in [0,\pi/5]\} \sqcup \{x'_\theta \, : \, \theta \in [0,\pi/5]\}\big].
\end{align*}
\end{theorem} The proofs of the complete characterization of both results are presented in \cref{proof:cop5,proof:cp5}. Note that in \cref{fig:circDNN-circCP-d5}, the region between the $\circDNN{5}$ and $\circCP{5}$ corresponds to quantum states that are PPT entangled. The two extremal $\circDNN{5}$ rays that are not elements of $\circCP{5}$ play an important role as \emph{extremal PPT entangled states}. 

\subsubsection{$d = 6$ and $d = 7$}

In these cases, the cones (due to symmetry) can be described by $4$ independent parameters. When we normalize the first parameter ($a_0 =1$), we have a convex set to describe in $3$ dimensions. Although we do not have the complete geometry of the $\circCP{d}$ cone, we characterize in this section the geometry on the three faces of this convex set with $a_i = 0$ for $i \in \{0,1,2\}$. To do this, let's define the face of the $\circCP{d}$ cone,  \[\circCP{d}^I = \{ x \in \circCP{d} \mid \supp{x} \subseteq I\} \] Looking at the slice of the $\circCP{6}$, and $\circCP{7}$ cones with $a_0 = 1$, there are $3$ free parameters that form a convex set. The next theorem completely characterizes these convex sets on each such face. The proofs can be found in \cref{sec:appendix-d-6-7}; we use that the fact that the $0$ in the vector restricts the possible supports of the terms $x * x^\rev$ in the decomposition, making it possible to characterize the duals completely. 

\begin{theorem}
Let $x_\theta :=(2 \cos{\theta}, 1,0,0,0,1)$. 
The extreme rays of the faces corresponding to a zero entry of $\circCP{6}$ are of the following form:
\begin{align*}
    \operatorname{ext} \circCP{6}^{\{0,1,2,4,5\}} &= \mathbb{R}_+\cdot\left[\{x_\theta * x_\theta^\rev \, : \,  \theta \in [0, \pi/3] \} \sqcup
    \left\{   (2,1,0,0,0,1), (1,0,1,0,1,0), \ket{0} \right\} \right]\\
    \operatorname{ext} \circCP{6}^{\{0,1,3,5\}} &= 
    \mathbb{R}_+\cdot\left\{ (2,1,0,0,0,1), (1,0,0,1,0,0), \ket{0} \right\}\\
    \operatorname{ext} \circCP{6}^{\{0,2,3,4\}} &= 
    \mathbb{R}_+\cdot\left\{ (1,0,1,0,1,0), (1,0,0,1,0,0), \ket{0} \right\}.
\end{align*}
\end{theorem}

\begin{figure}
    \centering
    \includegraphics[width=0.5\linewidth]{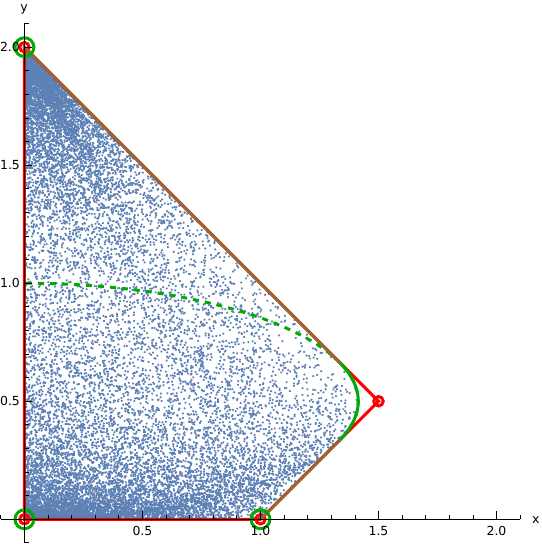}
    \caption{The slice $(2,x,y,0,y,x)$ through the cones $\circCP{6}$ and $\circDNN{6}$. In blue, randomly generated points inside the $\circCP{6}$ cone. In red, the four extreme points of $\circDNN{6}$ on this face, and the polytope they generate. The solid green curve is the continuous family of extremal points, while the four other green points are also extremal. The $\PPTcone$ entangled states are present in the small ``corner'' formed by the continuous family extremal rays of $\circCP{6}$ and the point $(2,3/2,1/2,0,1/2,3/2)$ (which is the only extremal $\circDNN{6}$ point that is not in $\circCP{6}).$}
    \label{fig:cp6-01245}
\end{figure}

From \cref{extremal-low}, we can conclude that the latter two the faces of the $\circCP{6}$ are equivalent to the faces of $\circDNN{6}$ as they have the same extremal rays. Hence, $\PPTcone$ entangled states are present only on the face $\circCP{6}^{\{0,1,2,4,5\}}$. In particular, $X_{a, a_0 \ket{e_0}, a}$ with $a = (4,3,1,0,1,3)$ is $\PPTcone$ entangled (and the only vector in $\operatorname{ext} \circDNN{6} \setminus \operatorname{ext}\circCP{6}$). In \cref{fig:cp6-01245}, we show this face of the cone with $a_0 = 2$. 

\bigskip

We now study the $\circCP{7}$ cone. We postpone the proof of the following theorem that characterises completely the facets of $\circCP{7}$ to the \cref{sec:appendix-d-6-7}. 

\begin{theorem}
Consider the following continuous families of vectors parametrized by a real parameter $\theta$:
\[
x^{(3)}_\theta := (2\cos(2\theta) + 4, 4\cos \theta, 1, 0, 0, 1, 4 \cos \theta),
\]
\[
x^{(1)}_\theta := (2\cos(2\theta) + 4, 0, 4\cos \theta, 1, 1, 4 \cos \theta, 0),
\]
\[
x^{(2)}_\theta := (2\cos(2\theta) + 4, 1, 0, 4\cos \theta, 4 \cos \theta, 0, 1).
\]
Then the faces of the $\circCP{7}$ cone can be completely characterized as:
\begin{align*}
\operatorname{ext} \circCP{7}^{\{0,1,2,5,6\}} &= \mathbb{R}_+ \cdot \left[
\left\{ x^{(3)}_\theta \, : \,  \theta \in [0, \pi/2] \right\} 
\sqcup \left\{ (2,1,0,0,0,0,1), \ket{0} \right\} 
\right], \\
\operatorname{ext} \circCP{7}^{\{0,2,3,4,5\}} &= \mathbb{R}_+ \cdot \left[
\left\{ x^{(1)}_\theta \, : \,  \theta \in [0, \pi/2] \right\} 
\sqcup \left\{ (2,0,1,0,0,1,0), \ket{0} \right\} 
\right], \\
\operatorname{ext} \circCP{7}^{\{0,1,3,4,6\}} &= \mathbb{R}_+ \cdot \left[
\left\{ x^{(2)}_\theta \, : \,  \theta \in [0, \pi/2] \right\} 
\sqcup \left\{ (2,0,0,1,1,0,0), \ket{0} \right\} 
\right].
\end{align*}
\end{theorem} 
\begin{figure}[htb]
    \centering
    \includegraphics[width=0.5\linewidth]{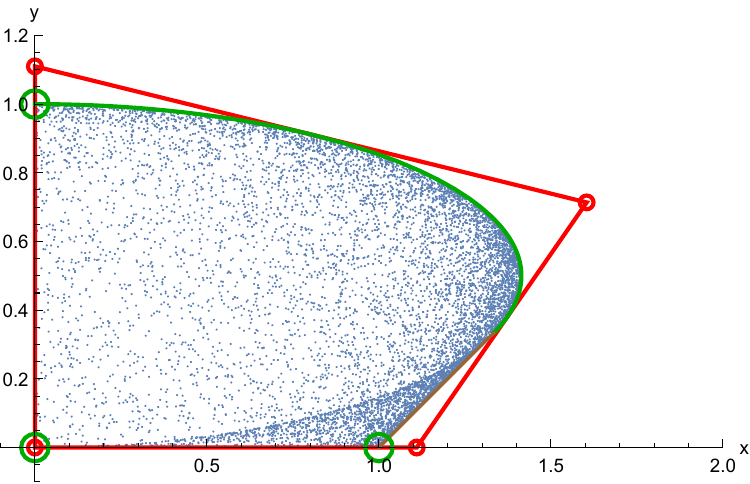}
    \caption{The slice $(2,x,y,0,0,y,x)$ through the cones $\circCP{7}$ (corresponding to separable matrices) and $\circDNN{7}$ (corresponding to $\PPTcone$ matrices). In blue, randomly generated points inside the face of the $\circCP{7}$ cone. In red, the four extreme points of $\circDNN{7}$ on this face, and the polytope they generate. The solid green curve is the continuous family of extremal points $x_\theta^{(3)}, \theta \in [0, \pi/2]$, while the three other green points (big green circles) are also extremal. All the extremal $\circDNN{d}$ states (red dots) except $\ket{0}$ are not in $\circCP{d}$. For $\circDNN{7}$ and $\circCP{7}$, all the other slices $(2,0,x,y,y,x,0)$ and $(2,y,0,x,x,0,y)$ are identical to the one above, up to permutation of the coordinates.}
    \label{fig:cp7-slices}
\end{figure}

\begin{figure}[htb]
    \centering
    \includegraphics[width=0.5\linewidth]{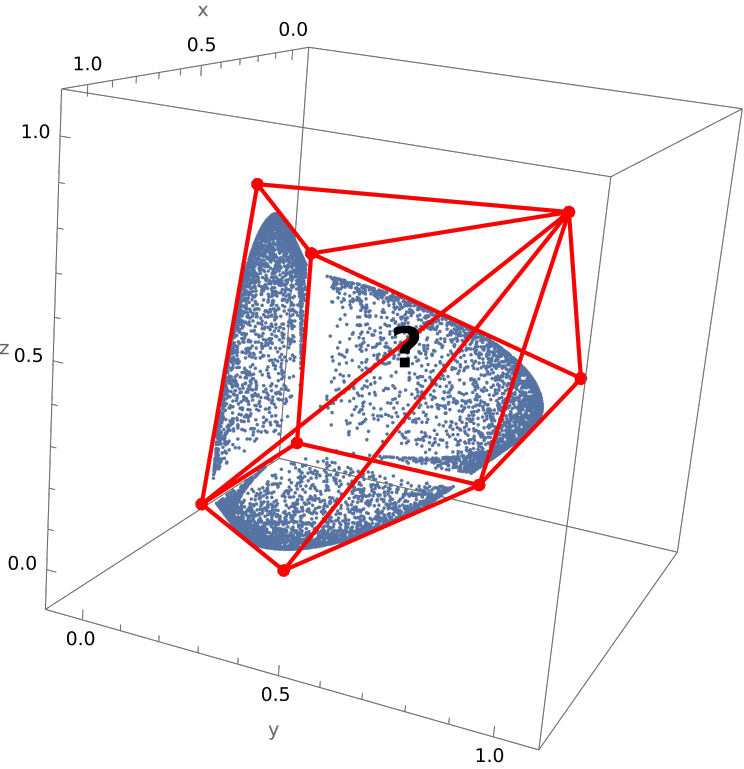}
    \caption{The figure shows the $(2,x,y,z,z,y,x)$ slice of the the $\circCP{7}$ and $\circDNN{7}$ cone. The polytope with the red edges/red vertices is the slice of $\circDNN{7}$. In blue, the faces with zeros of the $\circCP{7}$ cone and randomly generated points on the faces of $\circCP{7}$. The geometry of the bulk is an open question.}
    \label{fig:cp7-full}
\end{figure}

We display the slice of $\circCP{7}$ and $\circDNN{7}$ by setting $a_0 = 2$ (as $a_0 > 0$ for all non-zero elements of this cone) in \cref{fig:cp7-slices}. We leave the question of describing completely $\circCP{7}$ cone for future work. Analyzing the geometry of the $\circCP{}$ cones discussed in this section, we arrive at the following simple result (see also \cref{fig:cp7-full}).

\begin{theorem}
For every dimension $d$, and for the slice $a_0 =1$ there exists a ball of radius $\epsilon > 0$ around the point $\ket{e} = (1, 1, \ldots, 1)^\top \in \circCP{d}$ such that all $\circDNN{d}$ vectors in this ball are also in $\circCP{d}$.
\end{theorem}

\begin{proof}
Define the vector $\ket{x}_\epsilon = \ket{e} + (0, \epsilon, \epsilon, \ldots)^\top$. Then, we can do the following computation
\begin{align*}
    \ket{x}_\epsilon * \ket{x}_\epsilon^\rev &= \big(\ket{e} + (0, \epsilon, \epsilon, \ldots)^\top \big) * \big(\ket{e} + (0, \epsilon, \epsilon, \ldots)^\top \big)^\rev \\
    &= \ket{e} + \epsilon (d-1) \ket{e} + \epsilon^2 \big(d, (d-2), (d-2), \ldots, (d-2)\big)^\top,
\end{align*}

Rewriting this, we have:
\begin{align*}
    \ket{x}_\epsilon * \ket{x}_\epsilon^\rev = \big((d + (d-1)\epsilon) + d\epsilon^2\big) \ket{e} - \epsilon^2 \big(0, 2, 2, \ldots, 2\big)^\top.
\end{align*}

By definition, this vector belongs to $\circCP{d}$ for all $\epsilon \geq -1$. Let us define:
\[
    \tilde{\epsilon} := \frac{2\epsilon^2}{d + (d-1)\epsilon + d\epsilon^2} > 0, \quad \forall \epsilon > 0.
\]

Then, for any $x \in \circDNN{d}$ such that the Euclidean norm $\|x - \ket{e}\|_2 = \sqrt{\sum^{d-1}_{i = 1} |x_i - 1|^2} \leq \tilde{\epsilon}$, we can ensure that $x$ admits a $\circCP{d}$ decomposition.
\end{proof}

\begin{remark}
This result is analogous to the existence of a ball of separable matrices around the maximally mixed state $\rho = \mathbb{I}/d$ \cite{gurvits2002largest}. Determining the maximum radius of such a ball for the $\circCP{d}$, as a function of the local dimension $d$, remains an open question. Notice that the previous result cannot be obtained from the result in \cite{gurvits2002largest} because $X_{e/d,e_0,e/d} \neq \mathbb{I}/d$.
\end{remark}

We finally present in \cref{fig:cp7-full}, a partial geometry of the $\circCP{7}$ cone using the previous results, leaving the full case to be addressed in future work.

\subsection{Detecting \texorpdfstring{$\PPTcone$}{PPT} entangled mixtures of Dicke states} \label{ppt-entangled-states}

In this section, we address the question of detecting $\PPTcone$ entangled mixtures of Dicke states, which is equivalent to detecting if a matrix does not belong to the $\CP{}$ cone. Note that we address here the general (i.e.~not circulant) case, using techniques for the circulant case developed in this paper. This question has been addressed in the some of the previous papers relating entanglement and completely positive matrices, see e.g.~ \cite{tura2018separability}. We propose a new strategy to do this, by first projecting the matrices into the \emph{circulant} subspace, and then testing if they are in not in $\circCP{d}$. 

\begin{definition}
    We define the projection to the circulant subspace as $\mathcal{P}_d : \M{d} \rightarrow \mathsf{Circ}_{d} (\mathbb{C})$ as $X \mapsto \frac{1}{d} \sum^d_{k = 1} P^{-k} X P^k$. 
\end{definition}

It is easy to check that this operation preserves all the cones $\mathcal{P}_d (\CP{d}) \subseteq \CP{d}$, $\mathcal{P}_d (\DNN{d}) \subseteq \DNN{d}$ and also their duals $\mathcal{P}_d (\SPN{d}) \subseteq \SPN{d}$ and $\mathcal{P}_d (\COP{d}) \subseteq \COP{d}$. This allows us to conclude the next proposition, 

\begin{proposition}\label{prop:detect-not-CP-using-circulant-projection}
    Let $a \in \circDNN{d} \backslash \circCP{d}$. For any matrix $X \in \DNN{d}$ such that $\circulinv{\mathcal{P}_d(X)} = a$, we have that $X \notin \CP{d}$. Similarly for $a \in \circCOP{d} \backslash \circSPN{d}$, for any matrix $X \in \COP{d}$ such that $\circulinv{\mathcal{P}_d(X)} = a$, we have $X \notin \SPN{d}$. 
\end{proposition}

\begin{example}
In \cite{tura2018separability}, it is shown that the following $5 \times 5$ matrix is not in $\CP{5}$:
\[ A=
\begin{bmatrix}
    1 & 1 & 0 & 0 & 1 \\
    1 & 2 & 1 & 0 & 0 \\
    0 & 1 & 2 & 1 & 0 \\
    0 & 0 & 1 & 1 & 1 \\
    1 & 0 & 0 & 1 & 3
\end{bmatrix}.
\]

We can recover this result using \cref{prop:detect-not-CP-using-circulant-projection} as follows. First, project the matrix $A$ to the circulant subspace and extract the vector $a = (9,5,0,0,5)$. We claim that $a \notin \circCP{5}$. Indeed, since $a$ has a zero coordinate, it lies on the $x$ axis in \cref{fig:circDNN-circCP-d5}, with $x= 10/9$. We note that $1 < 10/9 < \sqrt 5-1$, proving the claim. Since $a \notin \circCP{5}$, we conclude by \cref{prop:detect-not-CP-using-circulant-projection} that $A \notin \CP{5}$. 
\end{example}

\section{A new class of PPT entangled states}\label{sec:examples}

In this section, we study the entanglement properties of the class of quantum states parametrized by two matrices $(A,C)$ such that $\operatorname{diag}(A) = \operatorname{diag}(C)$. \begin{equation*}
    X_{A,C} = \sum_{i \neq j} A_{ij} \ketbra{ij}{ij} + \ketbra{\omega}{\omega} +  \sum_{i \neq j} C_{ij} \ketbra{ij}{ji}
\end{equation*} where $\ketbra{\omega}{\omega}$ is the maximally entangled  (un-normalized) state $\sum_i \ket{ii}$. These states are exactly $\LDOIcone$ states introduced in the recent paper by the same authors, \cite{singh2021diagonal} but restricted to the matrix triple $(A, \mathbb{J}, C)$. All the convex properties of these states can be derived from the results about LDOI states \cite{singh2021diagonal}.

\begin{proposition}
    The following statements are true for $X_{A,C}$, 
    \begin{enumerate}
        \item $X_{A,C} \geq 0 \iff A_{ij} \geq |C_{ij}|^2, \, C= C^*$
        \item $X_{A,C} \in \PPTcone \iff A_{ij} \geq 1, \, C \geq 0$
    \end{enumerate}
\end{proposition} If $A$ and $C$ are circulant (also see \cref{ldoi-lcsi}) ,then the introduced state is local cyclic sign invariant, and can be parametrized by $a,c$ such that $\operatorname{circ}(a) = A$ and $\operatorname{circ} (c) = C$. We first look at the states with uniform diagonal, i.e $A = \mathbb{J}$ and show that the PPT and separability properties are characterized completely by the well-known correlation matrices. \cite{Christensen1979corr}.
\begin{theorem}
    In the case of $\operatorname{LDOI}$ matrices, we have: 
    \begin{itemize}
        \item $X^{\mathsf{LDOI}}_{(\mathbb{J},\mathbb{J}, C)} \in \PPTcone_d \iff C \in \mathsf{Corr}_d \coloneqq \{ Z\in \PSD{d} \, : \, \operatorname{diag}Z = \mathbb{I}_d \}$.
        \item $X^{\mathsf{LDOI}}_{(\mathbb{J},\mathbb{J}, C)} \in \SEPcone_d \iff C \in \operatorname{conv} \{ \ketbra{z}{z} : \ket{z}\in \mathbb{T}^d \} \subseteq \mathsf{Corr}_d$.
    \end{itemize}

    In particular, for local dimension $d \geq 4$,  there exist $\PPTcone$ entangled $\operatorname{LDOI}$ matrices with triples of the form $(\mathbb{J},\mathbb{J}, C)$. 
\end{theorem}

\begin{proof}
The first point was shown in \cite[see Example 3.6]{singh2021diagonal} and the discussion following it. For the second point, assume that $(\mathbb{J},\mathbb{J},C)$ is TCP so that it admits a decomposition given in Definition~\ref{def:PCP-TCP} with vectors $\ket{v_k} , \ket{w_k} \in \C{d}$. Since $B = \mathbb{J}$ is rank-$1$ and hence extremal in $\PSD{d}$, for all $k$,  $\ketbra{v_k \odot w_k}{v_k \odot w_k} \propto \ketbra{e}{e} \implies \ketbra{v_k \odot \overbar{w_k}} \propto \ketbra{z_k}$, where $\ket{z_k} \in \mathbb{T}^d$ is a phase vector. Thus, $$C\in \operatorname{conv} \{ \ketbra{z}{z} : \ket{z}\in \mathbb{T}^d \}.$$ However, for $d \geq 4$, there exist extreme points of $\mathsf{Corr}_d$ that are not rank one \cite{Grone1990corr, Loewy1980corr, Li1994corr}. Hence, there exist matrices $C\in \mathsf{Corr}_{d}$ that do not belong in $\operatorname{conv} \{ \ketbra{z}{z} : \ket{z}\in \mathbb{T}^d \}$ and for any such $C$, the triple $(\mathbb{J}, \mathbb{J}, C)$ cannot be TCP.
\end{proof}

This negatively answers the question posed in \cite[Proposition 3.6]{singh2021diagonal}

\begin{proposition}
\label{uniform-theorem}
    If $C$ is a circulant matrix, $C \in \{ \ketbra{z}{z} : \ket{z}\in \mathbb{T}^d \}$. Hence, the $\LCSIcone$ states with the vector triple $(\ket{e}, \ket{e}, c)$ are always separable if they are PPT. 
\end{proposition}

In a recent paper \cite{benkner2022characterizing}, the authors completely characterize the separable states in the set of LCSI states with $b := \ket{e}$ and $c = (1,0 \ldots 0)$. In the theorem that follows, we will prove a more general result in our framework that also reproduces, and also provides another proof of the results obtained in \cite{benkner2022characterizing}. Let us begin with with a simple lemma

\begin{lemma}
\label{lem:autocorrelation}
Let $x \in \C{d}$ be such that for all $l\in [d]$, $|\sum_i \overbar{x_i} x_{i-l}| = |\sum_i \overbar{x_i} x_{i}|$. Then, $x \in \mu \mathbb{T}^d$ for some $\mu\in \mathbb C$, i.e., for all $i\in [d]$, $|x_i|=|\mu|$.
\end{lemma} 
\begin{proof}
    We can rewrite the assumption of the lemma as 
    \begin{equation*}
        \forall l\in [d]: \qquad |\langle x | P^{-l} | x \rangle| = |\langle x | x \rangle |,
    \end{equation*}
    where $P$ is the shift permutation from Remark~\ref{Remark:shift}.
    Then, the equality condition of Cauchy Schwarz shows that $x$ is an eigenvector of $P$, i.e. $x$ is a scalar multiple of a phase vector. 
    \end{proof}

\begin{theorem}
\label{thm:extremal}
    Let $X_{a,b,c}$ be such that $a\in \R{d}_+$ such that $b \in \operatorname{ext} \circPSD{d}$ and $c \in \circPSD{d}$, and $c_0 := a_0$ (or vice-versa). Then, the following equivalences hold:
    \begin{alignat*}{2}
        a_l a_{d-l} \geq a_0^2 \quad \forall l \in [d]&&\iff X_{a,b,c} \in \PPTcone \\
        a_l\geq a_0  \quad \forall l \in [d]&&\iff X_{a,b,c} \in \SEPcone .
    \end{alignat*}
\end{theorem}

\begin{proof}
    The $\PPTcone$ equivalence follows easily from Theorem~\ref{theorem:circPPTPSD} (note that since $b$ is extremal, Remark~\ref{remark:extcircPSD} shows that $|b_i|=b_0=a_0$ for all $i$). Here, we prove the separable equivalence.
    Assume that $a_l \geq a_0\geq 0$ for all $l$. We split
    $$(a,b,c) = (a - a_0 \ket{e},0,0) + (a_0 \ket{e}, a_0 \ket{e} ,c), $$
    where the former triple is in $\circTCP{d}$ because $a_l - a_0 \geq 0$ for all $l$ and the latter triple is in $\circTCP{d}$ as $a_0 \ket{e}$ is a uniform vector and $\ket{e}, \mathcal{F}(c)\in \R{d}_+$, see \cref{uniform-theorem}. Hence, $X_{a,b,c}\in \SEPcone$ (note that in the forward implication, we did not actually make use of extremality of $b$, and this holds for all $b \in \circPSD{d}$).

    To show the converse, assume (wlog) that $b$ is extremal in $\circPSD{d}$ and $X_{a,b,c}\in \SEPcone$, so that $(a,b,c)$ admits a decomposition given in Definition~\ref{def:circTCP} with vectors $\ket{v_k}, \ket{w_k}$. Let $\ket{x_k}=\ket{v_k\odot w_k}$, so that we can write
    \begin{equation*}
        b = \sum_k \ket{\overbar{x_k}} * \ket{x_k}^\rev \implies \ket{\overbar{x_k}} * \ket{x_k}^\rev = \lambda_k \ket{b}
    \end{equation*}
    for some $\lambda_k\geq 0$, where the implication follows from the extremality of $b$. From Remark~\ref{remark:extcircPSD}, we know that $|b_0|=|b_i|$ for all $i$. Hence, we can use Lemma~\ref{lem:autocorrelation} to deduce that $\ket{x_k}\in \mu_k \mathbb{T}^d$ is a scalar multiple of a phase vector for each $k$, i.e., $|v^i_k w^i_k| = |\mu_k|$ for all $i,k$. Now, the $\circTCP{d}$ decomposition for $a$ shows that
   \begin{align*}
    \forall l\in [d]: \quad a_l = \sum_k \sum_i |v^i_k|^2 |w_k^{l-i}|^2 &= \sum_k |\mu_k|^2 \sum_i |v^i_k|^2 \frac{1}{|v^{l-i}_k|^2}  \\  
    &\geq d \sum_k |\mu_k|^2  \cdot \prod_i \left(|v^i_k|^2 \frac{1}{|v^{l-i}_k|^2} \right)^{1/d} \\ 
    &= d \sum_k |\mu_k|^2 \\ 
    &= a_0,
 \end{align*}
where we used the AM-GM inequality. This completes the proof.

\end{proof}

\begin{example}\label{ex:PPT-entangled-LCSI-3}
    We borrow the example from \cite[Example 9.1]{singh2021diagonal}: take $d=3$, $a := (2 \mu, 1, 4 \mu^2)$ and $b := (2 \mu, 2 \mu, 2 \mu)$ for $\mu \in \mathbb R$. The condition $b,c\in \circPSD{d}$ reads $\mu \geq 0$. 
    We have now that 
    \begin{alignat*}{2}
        X_{a,b,c} \in \PPTcone&\iff \mu \geq 0 \\
        X_{a,b,c} \in \SEPcone &\iff \mu = 1/2.
    \end{alignat*}
    Hence, the matrix $X_{a,b,c}$ is PPT entangled for any $\mu \in (0, \infty) \setminus \{1/2\}$. Such an example can be easily generalized to any dimension $d \geq 3$.
    \end{example}

\begin{remark}\label{rem:LCSI-PPT-entangled}
    \cref{thm:extremal} gives a very simple recipe to construct PPT entangled $\LCSI{}$ states in local dimension $d \geq 3$: take $b=c=e$, and $a$ such that
    $$a_1 = \alpha, \, a_{d-1} = 1/\alpha  \quad \text{ and } \quad a_i=1 \text{ for } i\neq 1,d-1$$
    for some $\alpha >1$. 
\end{remark}

We show the structure of these sparse $3 \times 3$ sparse PPT entangled states here. 
\[ X := 
\renewcommand{\arraystretch}{0.9}
\left(
\begin{array}{ccc|ccc|ccc}
\mu & 0      & 0      & 0      & \mu      & 0      & 0      &  0     &  \mu    \\
0      & \alpha & 0      & c      & 0      & 0      & 0      & 0      & 0      \\
0      & 0      & \mu/\alpha & 0      & 0      & 0      & \overbar{c}      & 0      & 0      \\ \hline
0      & \overbar{c}      & 0      & \mu/\alpha & 0      & 0      & 0      & 0      & 0      \\
\mu      & 0      & 0      & 0      & \mu & 0      & 0      &  0     & \mu      \\
0      & 0      & 0      & 0     & 0      & \alpha & 0      & c     &  0    \\ \hline
0      & 0      & c      & 0      & 0      & 0      & \alpha & 0      & 0      \\
0     & 0      & 0      & 0      &  0   & \overbar{c}      & 0      & \mu/\alpha & 0      \\
\mu      & 0      & 0      & 0      & \mu      & 0      & 0      & 0      & \mu \\
\end{array}
\right)
\]

If $\mathcal{F}(\mu, c, \overbar{c}) \geq 0$, then this state is separable if and only if $\alpha = \mu$ and entangled otherwise. 

\begin{remark}
If $b$ or $c$ in the above theorem is of the form $\beta \ket{e}+ (a_0 - \beta) \ket{0}$, the above argument doesn't hold, as it was essential to use the fact that the vectors $b$ is extremal. 
\end{remark}

\section{Conclusion and future directions}

We introduce and investigate bipartite mixed quantum states with local cyclic sign invariance. By leveraging the associated symmetry conditions, we show that these matrices can be parametrized in terms of triples of vectors. Exact conditions are derived for these vector triples to ensure that the corresponding matrices lie within the cones of positive semidefinite matrices and the $\PPTcone$ matrices. For vector triples, we define the concept of \emph{Circulant} Triplewise Complete Positivity ($\circTCP{}$), which provides a comprehensive characterization of separability. This framework enables the construction of simple examples of $\PPTcone$-entangled states in all dimensions $d \geq 3$. In the context of \emph{mixtures of Dicke states}, the $\PPTcone$ is shown to correspond to the semi-positive polyhedral cone of the Fourier matrix. We further establish new results regarding semi-positive cones and their supports, which may have independent significance for developing algorithms to enumerate extreme rays of these cones. One of the principal contributions of our work is the complete analytical characterization of the $\PPTcone$ and the set of separable states for $d \leq 5$ in mixtures of Dicke states with cyclic symmetry. Substantial progress is also achieved for the cases $d = 6$ and $d = 7$. Several examples of entangled mixtures of Dicke states available in the literature can be detected using the methods outlined in \cref{ppt-entangled-states}.

This work opens numerous avenues for future research. A key open problem is to understand the cone of circulant TCP vectors and to derive some better conditions for membership in this cone. This might be essential to provide new techniques to resolve the $\PPTcone^2$ conjecture for channels with cyclic sign covariance. The concept of factor width of PCP/TCP cones, introduced in \cite{singh2020ppt2}, has already been instrumental in proving the conjecture for $\mathsf{DUC}$ maps and is likely to play a significant role in addressing this problem.

Furthermore, it would be valuable to explore the entanglement properties of states invariant under other semi-direct product constructions with the diagonal orthogonal group. These additional symmetry constraints impose further structure on the matrix triples $(A, B, C)$ defining the LDOI states \cite{singh2021diagonal}. A particularly intriguing research avenue is to investigate the relationship between symmetry and $\PPTcone$ entanglement, particularly how much symmetry can be imposed on quantum states while ensuring the presence of $\PPTcone$ entanglement. As far as we are aware, the hyperoctahedral states \cite{park2024universal} are the most general class of invariant states for which the $\PPTcone$ condition implies separability, see \cref{tab:families} for reference. Is there even a larger class of states where this is true?

\bigskip

\noindent\textbf{Author Contributions. } All the authors contributed equally to this work. 

\bigskip

\noindent\textbf{Acknowledgments.} We would like to thank Sang-Jun Park for many insightful discussions and Jens Siewert for directing us to reference \cite{benkner2022characterizing}. A.G.~and I.N~were supported by the ANR project \href{https://esquisses.math.cnrs.fr/}{ESQuisses}, grant number ANR-20-CE47-0014-01. A.G is also supported by the EUR-MINT doctoral fellowship. S.S. is supported by the Cambridge Trust International Scholarship.

\bigskip

\noindent\textbf{Data availability.} Data sharing is not applicable to this article as no new data were created or analyzed in this study.

\appendix

\section{Circulant completely positive and copositive matrices}
\subsection{Extremal rays of polyhedral PPT cone}\label{sec:ext-vecDNN}

\begin{figure}[htb]
    \centering
    \includegraphics[width=0.4\linewidth]{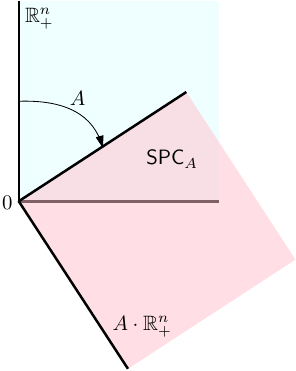}
    \caption{The semi-positive cone of a linear transformation $A: \R{n} \to \R{n}$ is the intersection of the non-negative orthant $\R{n}_+$ with its image through $A$.}
    \label{fig:semi-positive-cone}
\end{figure}

Since $\circDNN{d}$ is a polyhedral cone, it is generated by a finite number of extreme rays. As we shall see, the analytical enumeration of all extreme rays of $\circDNN{d}$ is still difficult problem. Here we prove certain results about this cone, and provide a simple algorithm to calculate its extreme rays. In this section, we shall continue using the correspondence between $\circDNN{d} \subseteq \R{d}$ and the real vector space $E_d \approx \R{n}$. We have 
$$\circDNN{d} \leftrightarrow \{b \in \R{n}_+ \, : \, G b \in \R{n}_+\}.$$
We are thus considering the so-called \emph{semi-positive cone} of the matrix $G$. We recall below the definition in the general case, see e.g.~\cite{sivakumar2018semipositive,hisabia2020properties} and references therein, as well as \cref{fig:semi-positive-cone}.

\begin{definition}
Let \(A\) be a real $n \times n$ square matrix. The \emph{semi-positive} cone with respect to \(A\) is the set
\[ \mathsf{SPC}_A := \{x \in \mathbb{R}^n_+ \, : \,  Ax \in \R{n}_+\}. \]
\end{definition}

Hence, finding the extremal rays of the cone $\circDNN{d}$ (in $\R{d}$) is equivalent to finding the extremal rays of the semi-positive cone of the matrix $G$ (in $\R{n}$). We shall focus on the latter problem in this section. For example, in the case $d=2$ (resp.~$d=3$) the matrix $G$ corresponds to a $\pi/4$ (resp. $\pi/6$) clock-wise rotation in $\R{2}$. The extremal rays of the semi-positive cone of $G$ are the first basis element $f_0$ and its image through $G$. We have the following result, showing that the extremal rays of $\mathsf{SPC}_G$ come in (possibly degenerate) pairs. 

\begin{proposition}\label{prop:x-Gx-dual-circ-DNN}
    If $x \in \R{n}_+$ is an extreme ray of $\mathsf{SPC}_G$ if and only if $Gx \in \R{n}_+$ is an extreme ray of $\mathsf{SPC}_G$.
\end{proposition}
\begin{proof}
    Since $G^2 = I$, $G$ leaves invariant its semi-positive cone $ G \cdot \mathsf{SPC}_G = \mathsf{SPC}_G$. Hence the set of extreme rays of $\mathsf{SPC}_G$ must have the same $G$-symmetry. Indeed, assume that x is an extreme ray of $\mathsf{SPC}_G$. Assume there exist vectors $g_1, g_2 \in \mathsf{SPC}_G$ such that $G(x) = g_1 + g_2$. Then $x =  G(g_1) + G(g_2)$. Since x is an extreme ray and $G(g_1), G(g_2) \in \mathsf{SPC}_G$, it $\implies G(g_1), G(g_2) \propto x \implies g_1, g_2 \propto G(x)$. Hence $G(x)$ is an extreme ray.
\end{proof}

The above proof shows that the extremal rays of $\mathsf{SPC}_G$ either satisfy $Ga = a$ (i.e.~they are eigenvectors of the matrix $G$ with eigenvalue $1$) or they come in pairs $(a,Ga)$. 

Let us now make some general observation about the semi-positive cones of real matrices. We shall specialize these results in the following subsection to the matrices $G$ corresponding to the $\circDNN{d}$ cones. The next observation is that the supports of the vectors defining the extreme rays are severely constrained. Recall that the support of a vector $v\in \C{n}$ is the set of indices of the non-zero elements of $v$:
\[ \supp(v) := \{ i \in [n] \,:\, v_i \neq 0\}. \]

We start with a slightly technical lemma. 
\begin{lemma}\label{lem:support-not-extremal}
  Let $X$ be a real $n \times n$ matrix and $a\in \mathsf{SPC}_X$. Assume that there exists a vector $b \in \R{d}$ such that the following conditions hold: 
  \begin{itemize}
      \item the vectors $a$ and $b$ are not colinear, i.e.~$\mathbb{R} a \neq \mathbb{R}b$
      \item $\supp(b) \subseteq \supp(a)$
      \item $\supp(Xb) \subseteq \supp(Xa)$
  \end{itemize}
  Then $\mathbb{R}_+ a$ is \emph{not} an extremal ray of $\mathsf{SPC}_X$.
\end{lemma}

\begin{proof}
    Put
    $$\epsilon := 
    \min \Bigg[ \min_{i \in \supp(a)} \Big(\frac{|a_i|}{|b_i|}\Big), \min_{i \in \supp(Xa)} \Big(\frac{|(Xa)_i|}{|(Xb)_i|}\Big) \Bigg] >0,$$
    where we use the convention $|x|/0 = +\infty$.
    Using the condition on the supports, we have $a \pm \epsilon b \in \mathsf{SPC}_X$, which provides a non-trivial decomposition of the ray $\mathbb{R}_+a$ inside the cone $\mathsf{SPC}_X$, proving the claim. 
\end{proof}

For the main result of this section, we shall denote by $X[J,I]$ the submatrix of $X$ consisting of rows indexed by $J$ and columns indexed by $I$ (here, $\emptyset \neq I,J \subseteq [n]$). We have 
$$X[J,I] : \R{I} \to \R{J}.$$

\begin{theorem}\label{thm:supports-cird-DNN}
Let $I$ and $J$ be subsets of $\{0, 1, \ldots, n-1\}$. Define the integer function 
$$\beta_X(I,J) := \dim \ker X[J^c, I].$$ 
We have then:
\begin{enumerate}
    \item If $\beta_X(I, J) \geq 2$ for some index sets $I$ and $J$, then any vector $a \in \mathsf{SPC}_X$ with $\supp(a) = I$ and $\supp(Xa) = J$ is not extremal in $\mathsf{SPC}_X$.
    
    \item If $\beta_X(I,J)= 1$ and $\ker X[J^c, I]$ is spanned by a vector $v$ such that $v >0$ entrywise and $X[J,I]v > 0$ entrywise, then the vector $a:=v_I \oplus O_{I^c}$ is extremal in $\mathsf{SPC}_X$ with $\supp(a)=I$, and $\supp(Xa) = J$.
    
    \item Conversely, let $a$ be a vector lying on an extremal ray of $\mathsf{SPC}_X$ with $\supp(a) = I$ and $\supp(Xa) = J$. Then $\beta_X(I,J) = 1$ and $\ker X[J^c, I]$ is spanned by $a_I > 0$ entrywise which also satisfies $(Xa)_J > 0$ entrywise.
\end{enumerate}

\end{theorem}

\begin{proof}

We will make repeated use of \cref{lem:support-not-extremal} to show this theorem. 
\begin{enumerate}
    \item Assume $\beta_X(I,J) \geq 2$ and let $a \in \mathsf{SPC}_X$ be a vector with $\supp(a) = I$ and $\supp(Xa) = J$. Since $\dim(\ker X[J^c, I])) \geq 2$, there exists at least one vector $b \in \ker X[J^c, I])$ not colinear to $a$. Hence, setting $b' := b_I \oplus 0_{I^c}$, we have 
    $$Xb' = X[J,I]b \oplus X[J^c,I]b = X[J,I]b \oplus 0_{J^c},$$
    thus $\supp(Xb') \subseteq J$. We can now apply \cref{lem:support-not-extremal}, proving the first claim.

    \item We only need to show extremality, all the other claims being clear. Assume, $a = b + c$ where $b,c \in \mathsf{SPC}_X$. We know that $\supp(b), \supp(c) \subseteq \supp(a) = I$ and similarly $\supp(Xb), \supp(Xc) \subseteq \supp(Xa) = J$, hence $b_I, c_I \in \ker X[J^c, I]$. Since the kernel
    has dimension $1$, $b_I$ and $c_I$ must be colinear with $v$ and thus $b,c$ must be colinear with $a$, proving the claim. 

    \item For an extremal vector $a$ with  $\supp(a) = I$ and $\supp(Xa) = J$, we have $X[J^c,I]a_I = 0$, hence $\beta_X(I, J) \geq 1$. Moreover, since $a$ is extremal, it follows from the first item in the result that $\beta_X(I, J) < 2 \implies \beta_X(I, J) = 1$. The strict positivity follows from the fact that $a \in \mathsf{SPC}_X$ and the support conditions.
\end{enumerate}

\end{proof}

The result above essentially tells us that, for every pair of subsets $\emptyset \neq I,J \subseteq [n]$, there is at most one extremal ray $a$ of $\mathsf{SPC}_X$ such that $\supp(a) = I$ and $\supp(Xa) = J$. Moreover, the pairs $(I,J)$ of supports of extremal rays have to satisfy $\beta_X(I,J) = 1$. Therefore the function $\beta_X(I,J)$ contains very useful information about the possible supports of the extremal rays of the cone $\mathsf{SPC}_X$.

\begin{lemma}
    The function $\beta_X$ has the following monotonicity properties with respect to the inclusion partial order on index sets: 
    \begin{align*}
        I \subseteq I' &\implies \beta_X(I,J) \leq \beta_X(I',J)\\
        J \subseteq J' &\implies \beta_X(I,J) \leq \beta_X(I,J').
    \end{align*}
\end{lemma}
\begin{proof}
    For the first point, assuming $I \subseteq I'$, if $v \in \ker X[J^c, I]$ then $v \oplus 0_{I'-I} \in \ker X[J^c, I']$, hence $\beta_X(I,J) \leq \beta_X(I',J)$. 

    For the second claim, let $v \in \ker X[J^c, I]$. Since $J'^c \subseteq J^c$, we have that $v \in \ker X[J'^c, I]$, proving the claim. 
\end{proof}

We have implemented a \texttt{Mathematica} routine to compute the extremal rays of $\mathsf{SPC}_X$ for an arbitrary matrix $X$ by enumerating the possible support sets, see \cite{semipositivecode}. For example, in the case of the matrix 
$$X = \begin{bmatrix}
    0 & 0 & 1 \\
    \tfrac 1 2 & \tfrac 1 5 & -1 \\
    0 & 1 &0
\end{bmatrix},$$
which was also considered in \cite[Example 3.3]{hisabia2020properties}, our code correctly identifies the extremal rays and their support: 

\begin{center}
\bgroup
\def\arraystretch{1.5}
  \begin{tabular}{|c|c|c|}
  \hline
  \rowcolor[HTML]{C0C0C0} 
  $\supp(a)$ & $\supp(Xa)$ & $a$ \\ \hline 
 $\{1\}$ & $\{2\}$ & $(1,0,0)$ \\ \hline
 $\{2\}$ & $\{2,3\}$ & $(0,1,0)$ \\\hline
 $\{1,3\}$ & $\{1\}$ & $\left(1,0,\frac{1}{2}\right)$ \\\hline
 $\{2,3\}$ & $\{1,3\}$ & $\left(0,1,\frac{1}{5}\right)$ \\\hline
\end{tabular}  
\egroup
\end{center}

\subsection{Enumeration of extremal rays}

\subsubsection{Facets of $\circDNN{d}$ for small $d$}\label{sec:facets-circDNN}

Using the \texttt{Mathematica} routine \cite{semipositivecode} we implemented for generating the extremal rays of semi-positive cones, we can generate the extremal rays of the doubly non-negative circulant cone $\circDNN{d}$ for small values of $d$, by first computing the extremal rays of $\mathsf{SPC}_G$ for the matrix $G$ and then embedding these vectors of $\R{n}$ into the larger space $\R{d}$ using the reverse basis change from \cref{eq:basis-change-F-G}; note the factor $\sqrt 2$ that has to be taken into account. Note how the two vectors
$$\ket 0 = (1, \underbrace{0,\ldots, 0}_{d-1 \text{ times}}) \quad \text{ and } \quad \ket{e} = (1,1,1,\ldots ,1)$$
are extremal for all $d \geq 2$. We present our results below, for $d=2,3,4,5,6$.

For $d=2$
\begin{center}
    \bgroup
    \def\arraystretch{1.5}
    \begin{tabular}{|c|c|c|}
      \hline
      \rowcolor[HTML]{C0C0C0} 
      $\supp(a)$ & $\supp(\mathcal{F} a)$ & $a$ \\ \hline 
     $\{1\}$ & $\{1,2\}$ & $(1,0)$ \\ \hline
     $\{1,2\}$ & $\{1\}$ & $(1,1)$ \\ \hline
    \end{tabular}
    \egroup

\end{center}

For $d=3$
\begin{center}
    \bgroup
    \def\arraystretch{1.5}
    \begin{tabular}{|c|c|c|}
      \hline
      \rowcolor[HTML]{C0C0C0} 
      $\supp(a)$ & $\supp(\mathcal{F} a)$ & $a$ \\ \hline 

$ \{1\}$ & $\{1,2,3\}$ & $(1,0,0)$ \\ \hline
 $\{1,2,3\}$ & $\{1\}$ & $\left(1,1,1\right)$ \\ \hline

    \end{tabular}
    \egroup
\end{center}

For $d=4$
\begin{center}
       \bgroup
    \def\arraystretch{1.5}
    \begin{tabular}{|c|c|c|}
      \hline
      \rowcolor[HTML]{C0C0C0} 
      $\supp(a)$ & $\supp(\mathcal{F} a)$ & $a$ \\ \hline 

$ \{1\}$ & $\{1,2,3,4\}$ & $(1,0,0,0)$ \\\hline
$ \{1,2,4\}$ & $\{1,2,4\}$ & $\left(1,\frac{1}{{2}},0,\frac{1}{{2}}\right)$ \\\hline
$ \{1,3\}$ & $\{1,3\}$ & $(1,0,1,0)$ \\\hline
$ \{1,2,3,4\}$ & $\{1\}$ & $\left(1,1,1,1\right)$ \\\hline

    \end{tabular}
    \egroup 
\end{center}

For $d=5$
\begin{center}
        \bgroup
    \def\arraystretch{1.5}
    \begin{tabular}{|c|c|c|}
      \hline
      \rowcolor[HTML]{C0C0C0} 
      $\supp(a)$ & $\supp(\mathcal{F} a)$ & $a$ \\ \hline 
 $\{1\}$ & $\{1,2,3,4,5\}$ & $\left(1,0,0,0,0\right)$ \\\hline 
 $\{1,2,5\}$ & $\{1,2,5\}$ & $\left(1,\frac{\sqrt 5 -1}{2},0,0,\frac{\sqrt 5 -1}{2}\right)$ \\\hline 
 $\{1,3,4\}$ & $\{1,3,4\}$ & $\left(1,0,\frac{\sqrt 5 -1}{2},\frac{\sqrt 5 -1}{2},0\right)$ \\\hline 
 $\{1,2,3,4,5\}$ & $\{1\}$ & $\left(1,1,1,1,1\right)$ \\\hline 

    \end{tabular}
    \egroup
\end{center}

For $d=6$
\begin{center}
    \bgroup
    \def\arraystretch{1.5}
    \begin{tabular}{|c|c|c|}
      \hline
      \rowcolor[HTML]{C0C0C0} 
      $\supp(a)$ & $\supp(\mathcal{F} a)$ & $a$ \\ \hline 

 $\{1\}$ & $\{1,2,3,4,5,6\}$ & $\left(1,0,0,0,0,0\right)$ \\\hline
 $\{1,2,6\}$ & $\{1,2,3,5,6\}$ & $\left(1,\frac{1}{{2}},0,0,0,\frac{1}{{2}}\right)$ \\\hline
 $\{1,3,5\}$ & $\{1,4\}$ & $\left(1,0,1,0,1,0\right)$ \\\hline
 $\{1,4\}$ & $\{1,3,5\}$ & $\left(1,0,0,1,0,0\right)$ \\\hline
 $\{1,2,3,5,6\}$ & $\{1,2,6\}$ & $\left(1,\frac{3}{4},\frac{1}{4},0,\frac{1}{4},\frac{3}{4}\right)$ \\\hline
 $\{1,2,3,4,5,6\}$ & $\{1\}$ & $\left(1,1,1,1,1,1\right)$ \\\hline

    \end{tabular}
    \egroup
\end{center}

\subsubsection{Analytical enumeration of facets of $\circDNN{d}$ with small supports}

Although the cone $\circDNN{d}$ has a polyhedral structure, the analytical enumeration of the extreme rays of this cone is still a significant challenge for general dimension $d$. In this section, we make some progress to understand the extreme rays of the $\circDNN{d}$ cone which have support of size $3$ or its Fourier transform (equal to the rank of the corresponding circulant matrix) has support of size $3$.

\begin{definition}
For every $I \subseteq [0 : d - 1]$, we define a facet of the cone $$\circDNN{d}^I= \mathrm{conv}\{a \in \mathrm{ext}(\circDNN{d}) \,:\, \mathrm{supp}(a) \subseteq I\}.$$
\end{definition}

Since extremal rays of facets are extremal in the cone, we have the following result. 

\begin{proposition}
    The extreme rays of the cone $\circDNN{d}^I$ are also extreme rays of the $\circDNN{d}$ cone.
\end{proposition}

\begin{lemma}\label{lem:scaling-lemma}
    For every subset $I = -I \subseteq [d]$ and any positive integer $k$, we have $\circDNN{md}^{mI} \cong \circDNN{d}^I$.  
\end{lemma}
\begin{proof}
The matrix $G$ satisfies the scaling property, $G_{mn}[mJ,I] = G_n[I,J]$. Since the extremal rays depend only on $\ker(G_n[J^c, I])$, we can show that the cones are isomorphic with $$a \in \circDNN{d}^I \longleftrightarrow b \in \circDNN{md}^{mI} \text{   such that   } b_{mi} = a_i \, \forall i \in [d] \text{  and  } b_k = 0 \text{  otherwise}.$$ 
\end{proof}

\begin{proposition}
    In the case of trivial support $I=\{0\}$, we have $\circDNN{d}^{\{0\}} = \operatorname{cone}\{ \ket 0\}$. 
    Hence both $\ket 0$ and $e$ are extremal rays of the $\circDNN{n}$ cone. Moreover these are the only extreme rays for $\circDNN{2}$ and $\circDNN{3}$. 
\end{proposition}

Since the analytical enumeration of the extremal rays is a difficult problem, we provide some partial results when the supports of the rays are small. We look at the facet $\circDNN{d}^{\{0,i, d-i\}}$ for each $i$ and $d$. 

\begin{proposition}
For $i \neq d/2$, the extreme rays of the $\circDNN{d}^I$ for $I = \{0, i, d-i\}$ are:
\label{extremal-low}
\begin{itemize}
    \item $\mathbb{R}_+ \ket{0}$
    \item 
    $\begin{cases}
        \mathbb{R}_+ \bigg( 2 \cos\left(\frac{\operatorname{gcd}(i,d) \pi}{d}\right), 1, 1 \bigg)_{\{0, i, d-i\}} & \text{if $\frac{d}{\operatorname{gcd}(i,d)}$ is odd}, \\
        \mathbb{R}_+ \bigg( 1, 1, 1 \bigg)_{\{0, i, d-i\}} & \text{if $\frac{d}{\operatorname{gcd}(i,d)}$ is even}.
    \end{cases}$

\end{itemize}
where $\operatorname{gcd}(p,q)$ denotes the greatest common divisor of two positive integers $p,q$. For $d$ even, and  $i = d/2$ the extreme rays with the support $I = \{0, d/2\}$ are $\mathbb{R}_+ (1,1)$ and $\mathbb{R}_+ (1,0)$. 
\end{proposition}

\begin{proof}
Let us start by considering the case $i \neq d/2$. The facet $\circDNN{d}^{\{0,i, d-i\}}$ is described by the inequalities $a_0 \geq 0$, $a_i=a_{d-i} \geq 0$,  and $a_0 + {2} \cos (\frac{2 \pi i j}{d}) a_i \geq 0$ for all $j$. Let $k=\operatorname{gcd}(i,d)$, and write $i=ka$ and $d=kb$. We have then
$$\min_{j \in [d]} \cos(\frac{2 \pi ij}{d})=\min_{j \in [d]} \cos(j \cdot \frac{2 \pi a}{b})=\min_{j \in [b]} \cos(\frac{2 \pi j}{b})=
\begin{cases}
    -1 &\quad \text{ if $b$ is even}\\
    -\cos\left( \frac{\pi}{b}\right) &\quad \text{ if $b$ is odd,}\\
    
\end{cases}
$$
where in the second equality above we have used the fact that $\exp(2 \pi \mathrm{i} \frac a b)$ is a \emph{primitive} root of unity. We conclude by plugging this minimum value in the inequality 
$$a_0/a_i \geq - {2} \min_{j \in [d]} \cos (\frac{2 \pi i j}{d}).$$

\medskip 
In the case of $d$ even, $i = d/2$, we can use the fact that $\circDNN{d}^{\{0, d/2\}} \cong \circDNN{2}^{\{0, 1\}} = \circDNN{2}$ by \cref{lem:scaling-lemma}.

\end{proof}

\begin{corollary}
The extremal rays of the $\circDNN{5}$ cone are exactly 
$\{\ket 0, e, a^{(1)}, a^{(2)}\}$, where
$$a^{(1)} = \left( 2 \cos(\pi/5), 1,0,0,1\right), \qquad  a^{(2)}=\left( 2 \cos(\pi/5), 0,1,1,0\right).$$
\end{corollary}

\begin{proof}
    These extremal rays have already been computed using computer assisted routine based on supports in \cref{sec:facets-circDNN}. We provide below a full analytical proof of this result. 
    By \cref{thm:supports-cird-DNN} we know that the unique extremal ray having full support is $e$; we get for free its dual $\ket 0$, see \cref{prop:x-Gx-dual-circ-DNN}. The only other possible supports are $\{0,1,4\}$ and $\{0,2,3\}$. These fall under \cref{extremal-low}, and we recover the vectors $a^{(1,2)}$ which can be rewritten as 
    $$a^{(1)} = \left(1, \frac{\sqrt 5 -1}{2},0,0,\frac{\sqrt 5 -1}{2}\right), \qquad  a^{(2)}=\left(1, 0,\frac{\sqrt 5 -1}{2},\frac{\sqrt 5 -1}{2},0\right),$$
    obtaining expressions that match the results from the previous subsection.
\end{proof}

\subsection{Extremal rays of the circulant $\text{COP}_5$ cone}
\label{proof:cop5}
\begin{theorem}\label{thm:circCOP-d5}
    Define, for $\theta \in \mathbb R$, the vectors
    \begin{align}
        h_\theta &:= (1, -\cos\theta, \cos(2\theta), \cos(2\theta), -\cos\theta)\\
        h'_\theta &:= (1,  \cos(2\theta), 
        -\cos\theta,
        -\cos\theta,
        \cos(2\theta)).
    \end{align}
    The extremal rays of the $\circCOP{5}$ cone are given by:
    $$
        \operatorname{ext} \circCOP{5} = \mathbb R_+\cdot \left[\{(0,1,0,0,1),(0,0,1,1,0) \}\sqcup\{ h_\theta : \theta \in [0,\pi/5]\} \sqcup \{h'_\theta : \theta \in [0,\pi/5]\}\right].
    $$
\end{theorem}
\begin{proof}
    Let us first show that the proposed rays are extremal. We start with the ray generated by the vector $(0,1,0,0,1)$, leaving the proof for the ray generated by $(0,0,1,1,0)$ to the reader. First, note that $(0,1,0,0,1) \in \circSPN{5} \subseteq \circCOP{5}$ since it is entrywise positive. Consider a decomposition 
    $$(0,1,0,0,1) = a+b, \qquad \text{with } a,b \in \circCOP{5}.$$
    Using \cref{lem:circCOP-necessary-positive}, we have $a_0, b_0 \geq 0$ hence $a_0 = b_0 = 0$. Moreover, taking $k=1$, we obtain $a_2,b_2 \geq 0$ and thus also $a_2 = b_2 = 0$. We conclude that the vectors $a$ is of the form $(0,a_1,0,0,a_1)$ and similarly for $b$, thus they are proportional to $(0,1,0,0,1)$, proving the extremality of the ray. 
    
    Let us now move on to the infinite families generated by the vectors $h_\theta$ and $h'_\theta$. Consider the characterization of the extremal rays of the copositive cone from \cite[Theorem 3.1]{hildebrand2012extreme}. Note that the first infinite family we propose correspond to the choice $T(\psi)$ from \cite{hildebrand2012extreme} with $\psi_i = \theta \in (0, \pi/5)$ for $i=1,2,3,4,5$. The value $\theta=0$ corresponds to the Horn matrix (which is extremal \cite{hall1963copositive}), while the value $\theta=\pi/5$ will be addressed later in the proof. The second family and the first family are conjugated by the (non-circulant!) permutation matrix 
    $$\begin{bmatrix}
        1 & 0 & 0 & 0 & 0\\
        0 & 0 & 1 & 0 & 0\\
        0 & 0 & 0 & 0 & 1\\
        0 & 1 & 0 & 0 & 0\\
        0 & 0 & 0 & 1 & 0
    \end{bmatrix},$$
    hence these are again extremal rays by \cite[Theorem 3.1]{hildebrand2012extreme}. 

    \medskip
    To finish the proof, we need to show that the proposed family are the only extremal rays. First, we claim that the only extremal rays $a$ with $a_0=0$ are the ones in the statement. Indeed, we have already shown that the slice $a_0=0$ of the $\circCOP{5}$ cone contains the extremal rays $(1,0)$ and $(0,1)$, in the $(x,y)$ parametrization of the $(0,x,y,y,x)$ slice. A cone in $\R{2}$ cannot have more than two extremal rays, proving the claim. To discuss extremal rays with $a_0 \neq 0$ (hence $a_0>0$ by \cref{lem:circCOP-necessary-positive}), we can restrict our attention on the slice $a_0=1$, see \cref{fig:circSPN-circCOP-d5}. Using \cref{lem:circCOP-necessary-positive}, we obtain $x=a_1=a_4 \geq -1$ and $y=a_2=a_3 \geq -1$. Hence, there are no elements of $\circCOP{5}$ (and thus no extreme points) below the $y=-1$ and to the left of the $x=-1$ lines in \cref{fig:circSPN-circCOP-d5}. Let us now show that $(1,-1,1,1,-1)$ (i.e.~the Horn point) is the only extremal point on the $x=-1$ line. This follows from the fact that any other point $(1,-1,y,y,-1)$, with $y>1$ can be decomposed as 
    $$(1,-1,y,y,-1) = (1,-1,1,1,-1) + (y-1)\cdot \underbrace{(0,0,1,1,0)}_{\in \operatorname{ext} \circCOP{5}},$$
    hence it cannot be extreme. 

    Consider now the fact that $(1,1,1,1,1) \in \circCP{5}$, as it can easily be seen by considering the convolution $e * e^\rev$. Hence
    $$(1,x,y,y,x) \in \circCOP{5} \implies \langle (1,1,1,1,1), (1,x,y,y,x) \rangle \geq 0 \iff x+y \geq - \frac 1 2.$$
    Graphically, this means that there are no (extreme) points of $\circCOP{5}$ strictly below the slanted red dashed line $x+y=-1/2$ in \cref{fig:circSPN-circCOP-d5}. Let us now consider the (extremal) points of $\circCOP{5}$ lying on this line. Clearly the two points $h_{\pi/5} = (1,-\cos(\pi/5), \cos(2\pi/5),\cos(2\pi/5),-\cos(\pi/5))$ and $h'_{\pi/5}=(1,\cos(2\pi/5), -\cos(\pi/5),-\cos(\pi/5),\cos(2\pi/5))$ are elements of $\circCOP{5}$ since they are (extremal) elements of $\circSPN{5}$. Note that they are the only elements of the families $h_\theta$, $h'_\theta$ lying on this line, so they must be extremal (the contrary would contradict the extremality of the other elements in the family); this proves the only remaining case from the beginning of the proof. Since they are extremal, no other points on the line $x+y=-1/2$ can be extremal, finishing the proof. 
\end{proof}

For $d=5$, as we only need three parameters to describe the cones of $\circCP{d}$ and $\circDNN{d}$, it is possible to visualize the complete cone after normalization. This visualization helps us gain more intuition about the set of separable as well as $\PPTcone$ entangled states in $d = 5$.

To do this, we look at the convex set of $\SEPcone$ states as the section obtained by setting $a_0 = 2$ in the $\circCP{5}$ cone. Essentially, the convex set we obtain provides us all the information as the any ray of the cone is $\lambda x$ where x is the extreme point of this set. From the last section, we know that the $\circCP{5}$ cone can be described as the intersection of all the half-planes parametrized by the parameter $\theta \in [0,\pi/5]$. In the next step, we explicitly calculate the extreme rays of the cone generated by these half planes. 

We shall need the following basic convexity lemma. 
\begin{lemma}\label{lem:convexity-extreme-point}
    Let $K \subseteq \R{2}$ be a convex set. Consider a $C^1$ family $(h_t)_{t \in (-1,1)}$ of extremal points of the dual, $h_t \in \ext K^\circ$, parametrized in a regular way such that 
    $$\delta_0 := \lim_{t \to 0} \frac{h_t - h_0}{t} \neq 0.$$
    Let $x_0 \in K$ be an element of $K$ lying on the supporting hyperplane defined by $h_0$: $\langle h_0, x_0 \rangle = 1$. Then $x_0$ is extremal in $K$: $x_0 \in \ext K$.
\end{lemma} 
\begin{proof}
    Assume that $x_0$ is not extremal in $K$, that is there exists $\Delta \in \R{2}$, $\Delta \neq 0$, such that $x_0 \pm  \Delta \in K$. First, note that $\Delta$ cannot be colinear to $x_0$:
    $$1 \geq \langle h_0, x_0 \pm \Delta \rangle = 1 \pm \langle h_0, \Delta \rangle \implies \langle h_0, \Delta \rangle = 0,$$
    while $\langle h_0, x_0 \rangle = 1$. Define, for $t \neq 0$, $\delta_t := (h_t - h_0)/t$. We have, for $t \neq 0$:
    $$1 \geq \langle h_t, x_0 \pm \Delta \rangle = 1 + t \langle \delta_t, x_0 \pm \Delta \rangle.$$
    Hence, we have
    \begin{align*}
        \forall t \in (-1,0) &\qquad \langle \delta_t, x_0 \pm \Delta \rangle \geq 0\\
        \forall t \in (0,1) &\qquad \langle \delta_t, x_0 \pm \Delta \rangle \leq 0.
    \end{align*}
    Taking directional limits $t \to 0^+$, respectively $t \to 0^-$, we obtain:
    $$\langle \delta_0, x_0 \pm \Delta \rangle = 0 \implies \langle \delta_0, x_0  \rangle = 0 \quad \text{ and } \quad \langle \delta_0, \Delta \rangle = 0.$$
    This, together with the fact that $x_0$ and $\Delta$ span $\R{2}$, contradicts the assumption $\delta_0 \neq 0$, finishing the proof.
\end{proof} 

\subsection{Extremal rays of the circulant $\text{CP}_5$ cone}

\label{proof:cp5}
\begin{proof}
    Nonzero elements $a$ of the $\circCP{5}$ cone must have $a_0 >0$, so it is enough to study the slice $a_0=2$ of this cone, see \cref{fig:circDNN-circCP-d5}. 

    Let us first show that the points in the statement are actually elements of $\circCP{5}$. We have 
    \begin{align*}
        (1,0,0,0,0) &= (1,0,0,0,0)*(1,0,0,0,0)^\rev\\
        (1,1,1,1,1) &= \frac 1 5 \cdot (1,1,1,1,1)*(1,1,1,1,1)^\rev\\
        (2,1,0,0,1) &= (0,0,1,1,0)*(0,0,1,1,0)^\rev\\
        (2,0,1,1,0) &= (0,1,0,0,1)*(0,1,0,0,1)^\rev.
    \end{align*}
    For the infinite families, write, for all $\theta \in [0,\pi/2]$, 
    $$x_\theta = v_\theta * v_\theta^\rev, \qquad \text{ with } v_\theta:=(2\cos \theta, 1, 0, 0, 1) \in \R{5}_+.$$
    Hence $x_\theta \in \circCP{5}$ for all $\theta \in [0,\pi/2] \supset [0,\pi/5]$; a similar result holds for $x'_\theta$. See the solid and dashed green curves in \cref{fig:circDNN-circCP-d5}.

    Let us now prove that the points in the statement are extremal. Since the coordinates of elements of $\circCP{5}$ are non-negative, $e_0$ is clearly extremal. Consider now the face 
    $$\{a \in \circCP{5} \, : \, \langle(0,1,0,0,1), a \rangle =0\}$$
    defined by the extremal point $(0,1,0,0,1)$ of the dual cone $\circCOP{5}$. Extremal points on this face must satisfy
    $$v_0v_4 + v_1 v_0 + v_2 v_1 + v_3v_2 + v_4 v_3 = 0, \qquad \text{ where } v_i \geq 0.$$
    Since we can shift cyclically the entries of the $v$ vector, assume $v_0 \neq 0$ $\implies$ $v_1=v_4=0$. Moreover, one of $v_{2,3}$ must be zero. Again, by cyclic permutation, we can assume $v_3=0$. Hence, this face consists of vectors
    $$(v_0,0,v_2,0,0)*(v_0,0,v_2,0,0)^\rev=(v_0^2+v_2^2, 0,v_0v_2, v_0v_2, 0)$$
    Thus it is equal to the set 
    $$\{(x,0,y,y,0) \, : \, 0 \leq 2y \leq x\}.$$
    The slice $x=2$ of this face is the segment $y \in [0,1]$ so it corresponds to the extremal rays $e_0$ and $(2,0,1,1,0)$. In a similar manner, considering the extremal hyperplane $(0,0,1,1,0) \in \operatorname{ext} \circCOP{5}$, one shows that $(2,1,0,0,1)$ is also an extremal ray of $\circCP{5}$. The ray supported on $(1,1,1,1,1)$ is an extremal element of $\circDNN{5}$ and an element of $\circCP{5}$ so it must be extremal in $\circCP{5}$. For the continuous families of points, observe that 
    $$\langle h_\alpha, x_\beta \rangle = 4(\cos \alpha - \cos \beta)^2.$$
    Hence, for all $\theta \in (0, \pi/5)$, we have 
    $$\forall t \in (0, \pi/5) \quad \langle h_t, x_\theta \rangle \geq 0 \qquad \text{ with equality iff } \qquad t = \theta.$$
    Note that the slice corresponding to setting the first coordinate of the vector to $1$ is a convex set of $\R{2}$, see \cref{fig:circSPN-circCOP-d5,fig:circDNN-circCP-d5}.  Thus, we can apply \cref{lem:convexity-extreme-point} to conclude that the point $x_\theta$ is extremal (in that slice), using the fact that the family of extremal hyperplanes $(h_t)_{t\in (0, \pi/5)}$ is $C^1$. The non-vanishing gradient condition is satisfied for the chosen parametrization: 
    $$ \nabla_t (-\cos t, \cos(2t)) = (\sin t, -2 \sin(2t) ) \neq 0 \qquad \forall t \in (0, \pi/5).$$

    Finally, let us prove that the points in the statement exhaust the set of extremal points of $\circCP{5}$. This follows, on the level of the slice $a_0=2$, from the analysis of the faces defined by the extremal rays of the dual cone $\circCOP{5}$, studied in \cref{thm:circCOP-d5}: 
    \begin{align*}
        (0,1,0,0,1) &\to \text{$1$-dim.~face }[x_0,(2,0,1,1,0)]\\
        (0,0,1,1,0) &\to \text{$1$-dim.~face }[x_0,(2,1,0,0,1)]\\
        h_0 &\to \text{$1$-dim.~face }[x_0,(2,0,1,1,0)]\\
        h'_0 &\to \text{$1$-dim.~face }[x'_0,(2,0,1,1,0)]\\
        h_{\pi/5} &\to \text{$1$-dim.~face }[x_{\pi/5},(2,2,2,2,2)]\\
        h'_{\pi/5} &\to \text{$1$-dim.~face }[x'_{\pi/5},(2,2,2,2,2)]\\
        h_\theta &\to \text{$0$-dim.~face } \{x_\theta\}\\
        h'_{\theta} &\to \text{$0$-dim.~face } \{x'_\theta\}.
    \end{align*}
\end{proof}

\subsection{Extremal rays for circulant $\CP{6}$ and $\CP{7}$ cones}\label{sec:appendix-d-6-7}

Recall that a circulant graph is a graph such that its adjacency matrix is circulant. A circulant graph with a connection set $I = - I$ on $d$ vertices has edges $i \leftrightarrow i \oplus s$ where $s \in I$.  We denote this graph by $C^I_d$. We do not consider graphs with loops so $0 \notin I$. We display in \cref{fig:C5-graph} the example of the circulant graph $C^{1,4}_5$. 
\begin{figure}[htb]
    \centering
    \includegraphics[width=0.2\linewidth]{graphs-C5.tex}
    \caption{A cycle on 5 vertices is the circulant graph $C^{1,4}_5$. The connection set $I=\{1,4\}$ signifies that the vertex $0$ is connected to vertices $1$ and $4=-1$.}
    \label{fig:C5-graph}
\end{figure}

A \emph{vertex cover} of a graph is a subset of vertices such that every edge is incident to at least one vertex of this set. A \emph{clique} of a graph is a subgraph that is complete. A maximal clique is a clique that is not contained in any strictly larger clique. Vertex covers correspond to cliques of the complimentary graph. For more details on graph properties, see \cite{bondy2008graph,diestel2024graph}. Using this, we prove an important lemma that uses the fact that when the support is not full, any element $x * x^R \in \circCP{d}$ will have only some allowed possible supports for $\operatorname{supp}(x)$. This approach is inspired by the results in \cite{berman2003completely} about completely positive matrices, and $\triangle$-free graphs. 

\begin{lemma}
\label{lem:support-lemma}
    For all terms $x * x^R \in \circCP{d}^I$, the $\operatorname{supp}(x)$ is a \emph{clique} in the circulant graph with the connection set $C^{I \backslash \{0\}}_d$. 
    Therefore, 
    $$\circCP{d}^I = \operatorname{cone}\{x * x^R : x \geq 0, \supp{x} \text{ is contained in a maximal clique of $C^{I \backslash \{0\}}_d$}\}$$
    Moreover, $x * x^R = P^lx * (P^lx)^\rev$ where $P^l = \circul{\ket{l}}$. 
\end{lemma} 

\begin{proof}
        We denote $I^c = [d-1] \backslash I$. Let us denote $a:=x * x^\rev \in \circCP{d}^I$. Since the support of $a$ is $I$, this means that means that $a_l = 0$ for all $l \in I^c$. Using the fact that these terms are non-negative, this means that $x_{i} x_{i + l} = 0$, so either $x_i = 0$ or $x_{i+l} = 0$ for all $l \in I^c$.
        Consider now $\circulant{I^c}{d}$, the circulant graph on $d$ vertices labeled as $[d]$ with the connection set $I^c$. Recall that the edges of this graph connect the points $i \leftrightarrow i+l$. Let there be some assignment of $0$ to $x_i$ such that $a_l = 0$ for $l \in I^c$. It will imply that there exists a vertex cover of the circulant graph where $x_i = 0$.  Recall that the complement of the vertex cover of graph is clique of the complementary graph. Moreover, we can show that if $\supp(x)$ is a clique of the graph $C^{I \backslash \{0\}}_d$, then $\supp(x * x^\rev) \in I$. In case of circulant graphs, $(C^{I^c}_d)^c = C^{I \backslash \{0\}}_d$. Therefore, the support (non-zero entries) of $x$ is \emph{any} \emph{clique} of the circulant graph $C^{I}_d$. Since all cliques are contained in maximal cliques, we are done. 
        
        The second statement follows from the fact that $P^l x * (P^l x)^\rev = \sum_i x_{i+l} x_{i+l-k} = \sum_i x_i x_{i-k} =  x * x^\rev$ for all $x$.    
    \end{proof}

Notice that since the cone $\circCP{d}^I$ does not have full support, the relint in the real vector space of symmetric vectors, $Q_d^\mathsf{sym} := \{a \in \mathbb{R} ^d\ \, : \,  a = a^\rev\}$ is empty. Since we will be computing the polar dual of this cone, we start by defining the the vector space that accurately captures the dimensions of these faces of the $\circCP{d}$ cone:
$$Q_I := \{x \in \R{d}: x = x^\rev \text{ such that } \operatorname{supp}(x) = I\}.$$ 
For any index set $I$, the dual of the cone $\circCP{d}^\circ$ will be computed in the space $Q_I$ and \emph{not} with respect to the vector space $Q_d^\mathsf{sym}$. We will use the following important result called the \emph{Fejér-Riesz theorem}, with important contributions by Szegő \cite{fejer1916trigonometrische,riesz2012functional}.
\begin{theorem}
\label{thm:fejer-riesz} 
    Let \( a = (a_0, a_1, a_2, \ldots, a_d) \) satisfy the following inequality for all \( \theta \): \[ \sum_{k=0}^d a_k \cos(k \theta) \geq 0. \] Then there exist coefficients \( c_0, c_1, \ldots, c_d \in \mathbb{R} \) such that \[ a_k = \sum_{j=0}^{d-k} c_j c_{j+k}. \] 
\end{theorem}

Define the following continuous families of vectors:
\begin{align*}
    h^{(3)}_\theta &:= (1, -\cos \theta,\cos(2\theta),0,\cos(2\theta),-\cos\theta)\\
    x_\theta &:= (2 \cos \theta, 1,0,0,0,1).
\end{align*}
We can state now the main result of this appendix, regarding the dual of the face of the $\circCP{6}$ cone given by setting to zero the fourth coordinate. 

\begin{theorem}
The extremal rays the cone $(\circCP{6}^I)^\circ$, corresponding to the face given by $I = \{0,1,2,4,5\}$, are:
\begin{itemize}
    \item $\{\mathbb{R}_+ h^{(3)}_\theta \}_{\theta \in [0, \pi/3]}$
    \item $\mathbb{R}_+(0,0,1,0,1,0)$
    \item $\mathbb{R}_+ (0,1,0,0,0,1)$
\end{itemize}
\end{theorem}

\begin{proof}
We first begin by analyzing the graph $C_6^I$ (see \cref{fig:c6-12}) for $I = \{1,2,4,5\}$. We observe that all the maximal cliques of this graph are of the form $\{0,1,2\} \oplus p_1$ and $\{0,2,4\} \oplus p_2$ for $p_1, p_2 \in \{0,1,\ldots,5\}$.

\begin{figure}[H]
    \centering
    \includegraphics[width=0.2\linewidth]{graphs-C6.tex}
    \caption{The graph $C_6^{1,2}$}
    \label{fig:c6-12}
\end{figure}
By defining 

\[K_1 := {\cone\{x * x^\rev \,:\,  x \geq 0, \, \supp(x) \subseteq \{0,1,2\}\}}\] 
\[K_2  := {\cone\{x * x^\rev \,:\, x \geq 0, \, \supp(x) \subseteq \{0,2,4\}\}}\] 
we can use \cref{lem:support-lemma} to show that,  \(\circCP{6}^I = K_1 + K_2\). Using the results about convex cones in \cref{thm:dual-cone-sum}, it follows that $(\circCP{6}^I)^\circ = K_1^\circ \cap K_2^\circ$. We claim that the following is true 
\begin{equation} \label{eqn:k1} K_1 = \cone\{a \in Q_I \, : \, a \geq 0, \, \langle h^{(3)}_\theta, a \rangle \geq 0 \quad \forall \theta \in [0, \pi)\}. \end{equation} 
This is easy to see. The elements of $K_1$ are of the form $x * x^\rev$ for $x = (c_0, c_1, c_2,0,0,0)$ (which has the support \{0,1,2\}) for $c_0, c_1, c_2 \geq 0$. 
\[
a:= x * x^\rev = (c_0^2 + c_1^2 + c_2^2, c_0 c_1 + c_1 c_2, c_0 c_2, 0, c_0 c_2, c_0 c_1 + c_1 c_2). \]

Each such vector satisfies the inequality, 
\begin{align*}
 \langle h^{(3)}_\theta, a \rangle 
&= c_0^2 + c_1^2 + c_2^2 - 2 (c_0 c_1 + c_1 c_2)\cos\theta  + 2 c_0 c_2\cos(2\theta)  \\
&= \operatorname{Re}\big|c_0 - c_1 e^{\mathrm{i}\theta} + c_2 e^{\mathrm{i}2\theta}\big|^2 \geq 0
\end{align*}
and elements $x * x^\rev$ are non-negative for $x \geq 0$. This is also true for any element $K_1$, which is just a sum of elements of this form. Now assume that some vector $a \in Q_I$ that satisfies $a \geq 0$ and also the continuous family of inequalities
\[
\forall \theta \in [0,\pi), \qquad \langle (1, -\cos\theta, \cos(2\theta), 0, \cos(2\theta), -\cos\theta) , a \rangle \geq 0.
\]
By the Fejér-Riesz \cref{thm:fejer-riesz}, we can write $a = (c_0,c_1,c_2,0,0,0) * (c_0,c_1,c_2,0,0,0)^\rev$ for $c_i \in \mathbb{R}$.
Since $a \geq 0$, we have that all the parameters $c_i \geq 0$ or $c_i \leq 0$. In the first case, it shows that $a \in K_1$. In the latter case, we can flip all the signs to show that $a = c * c^\rev = (-c) * (-c)^\rev \in K_1$, proving the claim in \cref{eqn:k1}.

\medskip

\cref{eqn:k1} implies that the dual cone (remember that the duals are taken in $Q_I$) of $K_1$ is given by the inequalities defining $K_1$ itself:
\[
K_1^\circ = \operatorname{cone}\left\{\{h^{(3)}_\theta\}_{\theta \in [0, \pi)} \sqcup \{(0,1,0,0,0,1), (0,0,1,0,1,0)\}\right\}.
\]

For $\theta \in (0,\pi/2)$, note that 
$$h^{(3)}_{\pi - \theta} = h^{(3)}_\theta + 2 \cos \theta \cdot (0,1,0,0,0,1).$$
We now claim that (see \cref{fig:cp-6-face}, left panel, black curve)
\[
\ext K_1^\circ = \operatorname{cone}\left\{\{h^{(3)}_\theta\}_{\theta \in [0, \pi/2]} \sqcup \{(0,1,0,0,0,1), (0,0,1,0,1,0)\}\right\}.
\]
The ``$\subseteq$'' inclusion is clear. Fix $\theta \in [0,\pi/2)$. To show that $h^{(3)}_\theta$ is extremal, assume $h^{(3)}_\theta = \sum \lambda_i h^{(3)}_{\theta_i} + \mu_1 (0,1,0,0,0,1) + \mu_2 (0,0,1,0,1,0)$ with $\lambda_i, \mu_j > 0$ and $\theta_i \in [0,\pi/2]$. Considering the scalar product with 
$$x_\theta * x_\theta^\rev = (2+4\cos^2\theta, 4 \cos \theta, 1,0,1, 4 \cos \theta),$$
we get:
\begin{align*}
    0 &= \langle h^{(3)}_{\theta}, x_\theta * x_\theta^\rev \rangle \\
    &= \sum_i \lambda_i \langle h^{(3)}_{\theta_i} , x_\theta * x_\theta^\rev \rangle + \mu_1 \langle (0,1,0,0,0,1), x_\theta * x_\theta^\rev \rangle + \mu_2 \langle  (0,0,1,0,1,0), x_\theta * x_\theta^\rev \rangle\\
    &= \sum_i \lambda_i \cdot 4(\cos\theta_i - \cos\theta)^2 + \mu_1 \cdot 8 \cos \theta + \mu_2 \cdot 2.
\end{align*}
This implies $\theta_i = \theta$ for all $i$ and $\mu_2 = 0$; the case $\theta = \pi/2$ can be easily dealt with separately. Thus, $h^{(3)}_\theta$ is extremal. The extremality of the rays generated by the vectors $(0,1,0,0,0,1)$ and $(0,0,1,0,1,0)$ follows easily from their zero-pattern.

\bigskip

Let us analyze now the cone $K_2$ and its dual. The elements of $K_2$ are of the form $x * x^\rev$ for $x = (c_0, 0, c_1, 0, c_2, 0)$ with $c_{0,1,2} \geq 0$. Expanding this, we have
\[a:= x * x^\rev = (c_0^2 + c_1^2 + c_2^2, 0, c_0 c_1 + c_1 c_2 + c_2 c_0, 0, c_0 c_1 + c_1 c_2 + c_2 c_0, 0).\]
Such vectors $a$ are of the form $a=(a_0, 0, a_2, 0, a_2, 0)$ and satisfy $a_{0,2} \geq 0$ and $a_0 \geq a_2$. Conversely, it is easy to see that any such vector $a$ satisfying these inequalities can be factorized as $a:= x * x^\rev$ with $x \in Q_I$, $x \geq 0$. Therefore, $K_2 = \cone\{a \in Q_I \, : \, a \geq 0 \text{ and } a_0 \geq a_2\}$ hence its dual (in $Q_I$, which is the vector space of $b \in \R{6}$ with $b_3=0$) is easily computed: 
\[K_2^\circ = \operatorname{cone}\left\{\{(1,x,-1/2,0,-1/2,x) : x \in \mathbb{R}\} \cup \{(0,1,0,0,0,1),(0,0,1,0,1,0\} \right\}\]
Note that support of $K_2$ is a strict subset of the support of $K_1$, and the dual of $K_2$ is \emph{not} pointed. Putting everything together, we arrive at the following result,
\[
\ext(\circCP{6}^I)^\circ = \ext(K_1^\circ \cap K_2^\circ) = \Big\{\mathbb{R}_+ h^{(3)}_\theta \Big\}_{\theta \in [0, \pi/3]} \sqcup \{\mathbb{R}_+(0,1,0,0,0,1), \mathbb{R}_+(0,0,1,0,1,0)\}, 
\]
hence completing the proof. We plot in \cref{fig:cp-6-dual} the illustration of the geometry of the slice $(1,x,y,0,y,x)$ of this cone. 
\end{proof}

\begin{theorem}
The extremal rays the cone $\circCP{6}^I$, corresponding to the face given by $I = \{0,1,2,4,5\}$, are:
\begin{itemize}
    \item $(1,0,1,0,1,0)$
    \item $(1,0,0,0,0,0)$
    \item $(2,1,0,0,0,1)$.
    \item $x_\theta * x_\theta^\rev$ for $\theta \in [0, \pi/3]$
\end{itemize}
\end{theorem}

\begin{proof}
    The first three vectors are extremal as they belong to $ \operatorname{ext}\circDNN{6}$ and have the following $\circCP{6}$ decompositions
    $$(1,0,1,0,1,0) = 1/3 \cdot (1,0,1,0,1,0) * (1,0,1,0,1,0)^\rev$$
    $$(1,0,0,0,0,0) = (1,0,0,0,0,0) * (1,0,0,0,0,0)^\rev$$
    $$(2,1,0,0,0,1) = (1,1,0,0,0,0) * (1,1,0,0,0,0)^\rev$$
    For the continuous families of points, observe that $$\langle h^{(3)}_\alpha, x_\beta \rangle = 4(\cos \alpha - \cos \beta)^2$$
    Hence, for all $\theta \in (0, \pi/3)$, we have 
    $$\forall t \in (0, \pi/3) \quad \langle h^{(3)}_t, x_\theta \rangle \geq 0 \qquad \text{ with equality iff } \qquad t = \theta.$$
    Note that the slice corresponding to setting the first coordinate of the vector to $1$ is a convex set of $\R{2}$, see \cref{fig:cp6-01245}.  Thus, we can apply \cref{lem:convexity-extreme-point} to conclude that the point $x_\theta$ is extremal (in that slice), using the fact that the family of extremal hyperplanes $(h^{(3)}_t)_{t\in (0, \pi/3)}$ is $C^1$. The non-vanishing gradient condition is satisfied for the chosen parametrization: 
    $$ \nabla_t (-\cos t, \cos(2t)) = (\sin t, -2 \sin(2t) ) \neq 0 \qquad \forall t \in (0, \pi/3).$$   
\end{proof}

\begin{figure}[ht]
    \centering
    \includegraphics[width=0.6\textwidth]{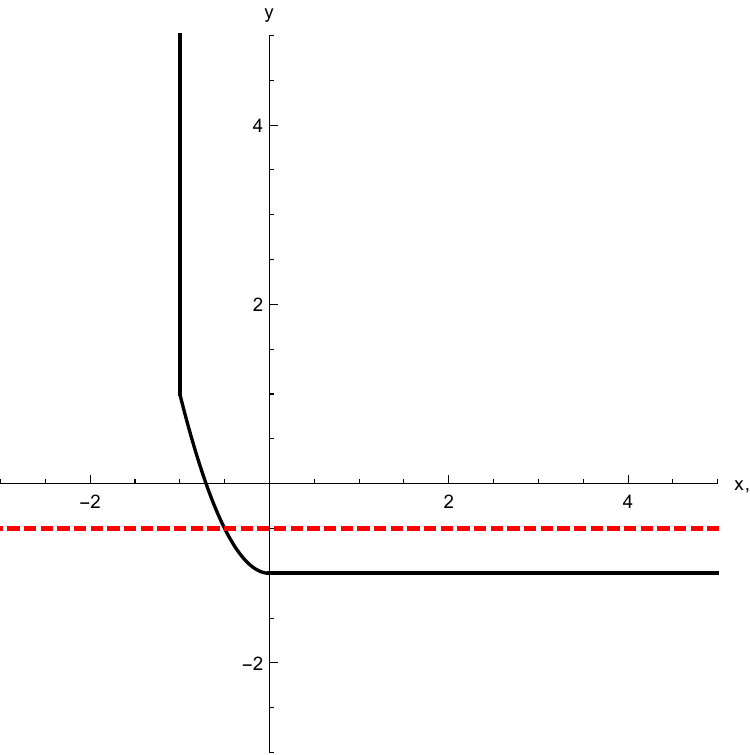}
    \caption{The individual boundaries of the slices of $K_1^\circ$ (black) and $K_2^\circ$ (red, dashed).}
    \label{fig:cp-6-face}
\end{figure}

\begin{figure}[ht]
    \centering
    \includegraphics[width=0.6\textwidth]{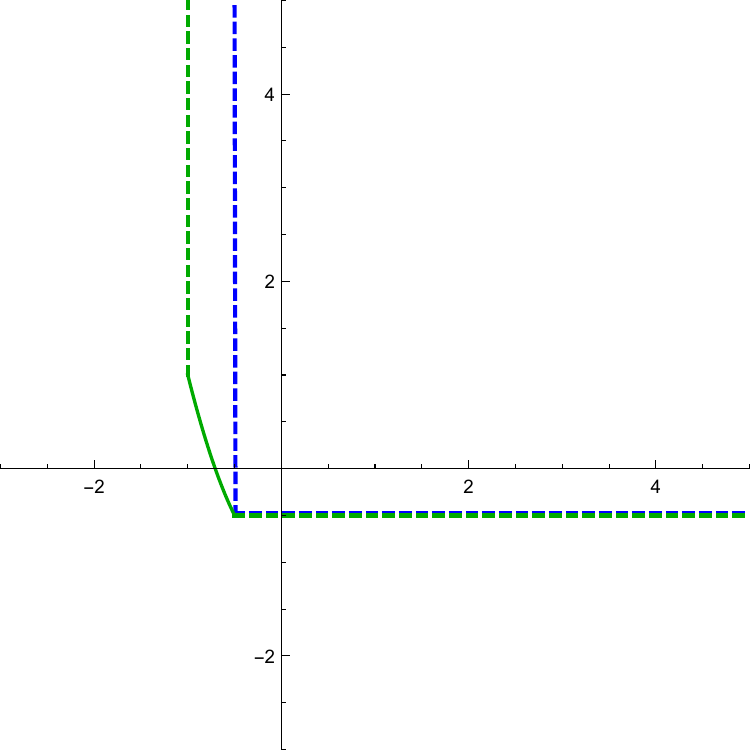}
    \caption{The intersection of the slices (green boundary) and its relation to the boundary of $\circSPN{6}$ (blue).}
    \label{fig:cp-6-dual}
\end{figure}

Similar computations can be done for the faces of the cone  $d=7$. We provide the basic steps needed for characterizing the geometry of the face $\circCP{7}^I$ for $I = \{0,1,2,5,6\}$. Again, we proceed by looking at the maximal cliques of the graph $C^{\{1,2\}}_7$ (as in \cref{fig:graph-c7-12}) and using the \cref{lem:support-lemma}. 

\begin{figure}[H]
    \centering
    \includegraphics[width=0.2\linewidth]{graphs-C7.tex}
    \caption{The graph $C^{\{1,2\}}_7$}
    \label{fig:graph-c7-12}
\end{figure}

\begin{figure}
    \centering
    \includegraphics[width=0.5\linewidth]{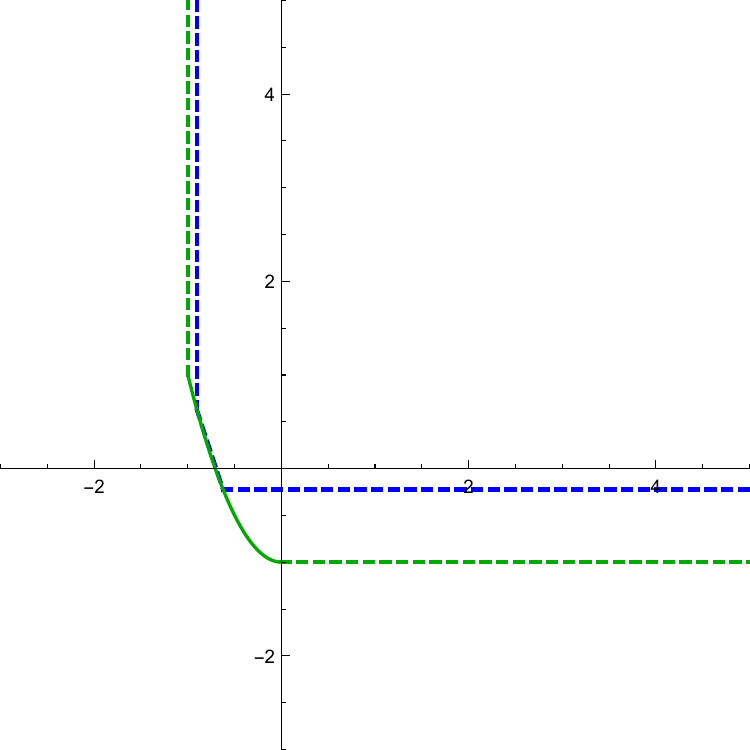}
    \caption{The following figure describes the slice $(1,x,y,0,0,y,x)$ of the $\circCP{7}^\circ$ cone. In green, we have the boundary of the cone $\circCP{7}^\circ$. In blue, we have the cone $\circSPN{7}$ which is the dual of the cone $\circDNN{7}$.}
    \label{fig:cp-7-dual}
\end{figure}

The maximal cliques of the graph are $\{0,1,2\} \oplus p$ for all $p \in \{0,1,2,3,4,5\}$. Using the same arguments as for the $\circCP{6}$ cone, one can show that the dual cone (w.r.t.~the space $Q_I$) is completely characterized as (see \cref{fig:cp-7-dual}) $$\operatorname{ext} (\circCP{7}^I)^\circ = \Big\{\mathbb{R}_+h^{(3,4)}_\theta\Big\}_{\theta \in [0,  \pi/2]} \sqcup \{\mathbb{R}_+(0,1,0,0,0,0,1), \mathbb{R}_+(0,0,1,0,0,1,0)\}$$
where $h^{(3,4)}_\theta := (1,\cos \theta, \cos(2\theta), 0,0, \cos(2\theta), \cos \theta)$.
This allows to completely determine the face of the cone $\circCP{7}$. By defining $v_\theta := (2 \cos \theta,1,0,0,0,0,1)$ it can be shown that for $I = \{0,1,2,5,6\}$ 

$${\operatorname{ext}{\circCP{7}^I}} = \Big\{\mathbb{R}_+(v_\theta * v_\theta^\rev)\Big\}_{\theta \in [0,\pi/2]} \sqcup \{\mathbb{R}_+\ket{0}, \mathbb{R}_+(2,1,0,0,0,0,1)\}$$ which is shown in the \cref{fig:cp7-slices}. Moreover, the following graphs are isomorphic $C_7^{\{1,2\}} \cong C_7^{\{2,3\}} \cong C_7^{\{3,1\}}$, allowing for the same computations to be done for all faces. 

\bibliography{references}
\bibliographystyle{unsrt}
\bigskip
\hrule
\bigskip

\end{document}